\def\D{{\cal D}}
\def\Loc{L_{\rm loc}}
\def\kbar{\overline{K}}
\def\figwidth{9cm}
\documentclass{report}
\usepackage{svcon2e}

\begin{document}

\pagenumbering{arabic}
\chapter{Disordered Quantum Solids}
\chapterauthors{T. Giamarchi, E. Orignac}

\begin{abstract}
Due to the peculiar non-fermi liquid of one dimensional systems,
disorder has particularly strong effects. We show that such
systems belong to the more general class of disordered quantum
solids. We discuss the physics of such disordered interacting
systems and the methods that allows to treat them. In addition
to, by now standard renormalization group methods, We explain how
a simple variational approach allows to treat these problems even
in case when the RG fails. We discuss various physical
realizations of such disordered quantum solids both in one and
higher dimensions (Wigner crystal, Bose glass). We investigate in
details the interesting example of a disordered Mott insulator and
argue that intermediate disorder can lead to a novel phase, the
Mott glass, intermediate between a Mott and and Anderson
insulator.
\end{abstract}

\section{Introduction}
Disorder effects are omnipresent in condensed matter physics,
where one has to struggle very hard to deal with clean systems.
Quite remarkably fermionic system exhibit marked differences with
disorder in classical systems. Indeed the very existence of a
Fermi energy $E_F$ ``reduces'' the effects of disorder since the
relevant parameter now become the relative strength of the
disorder compared to the Fermi energy $D/E_F$ or the mean free
path compared to the Fermi length $k_F l$. Of course nature would
not remain as simple as that, and quantum effects lead in fact
also to reinforcement of the disorder effects and turn in low
dimension a free electron system into an insulator, as pointed
out by Mott and Twose\cite{mott_loc}.

We have now gained a very good understanding of the properties of
such disordered free electron systems. To tackle them an arsenal
of methods ranging from
diagrams\cite{berezinskii_conductivity_log,abrikosov_rhyzkin},
scaling theory\cite{abrahams_loc},
replicas\cite{wegner_localisation,efetov_localisation},
supersymmetry\cite{efetov_supersym_revue} have been developped.
Life become much less simple when interactions among fermions are
taken into account. Although it is very intuitive to think that
when interactions are small the noninteracting problem is a good
starting point, such an intuition turns out to be wrong for a
number of reasons: (i) even if in the pure system interactions
can be ``removed'' from the system by resorting to Fermi Liquid
theory\cite{nozieres_book} this is not the case when disorder is
present. Because disorder renders electrons  slowly diffusive
slowly rather than ballistic, they feel the interactions much
more strongly with explosive results\cite{altshuler_aronov}.
Effective interactions increase when looking at low energy
properties and Fermi liquid theory breaks down. The
non-interacting physics is thus only relevant for very low
disorder. (ii) When the dimension is small or the interactions
strong to start with (like in systems undergoing Mott
transitions) it is of course impossible to start from the
non-interacting limit and one has to solve the full problem. What
puts us at a disadvantage here is that most of the techniques
useful for the noninteracting case also fail as soon as
interactions are included: (i) supersymmetric method, which rests
by construction on the quadratic nature of the hamiltonian is
useless. (ii) it is now difficult now to really determine classes
of diagram to sum.

Disorder can still be averaged over by using replicas but then
one is left with a complicated (and untractable) theory.
Renormalisation group attempts have been made with some
success\cite{finkelstein_localization_interactions,lee_mit_long}
but also with the problem that the coupling constants diverge, so
that the low energy fixed point remains elusive. Many other
attempts in treating this very complicated problem exist in the
literature and it is impossible to list them
all\cite{belitz_localization_review}. Note that numerical studies
are also hampered when studying this problem both because of the
difficulty in taking the interactions into account (which more or
less imposes either Monte-Carlo or exact diagonalization) and then
to perform the complicated disorder averaging with enough
statistics or large system sizes to get reliable results.

A very peculiar situation occurs when one considers one dimension.
On one hand one expects the difficulty to be maximum here. The
interactions lead to very strong effects and destroy any trace of
Fermi liquid giving rise to what is known as a Lutinger liquid.
The disorder is also extremely strong, giving rise for the
noninteracting case to a system so localized that the
localisation length is simply the mean free path and a diffusive
regime is absent. On the other hand one is in $d=1$ in a much
better situation to tackle the problem since the interactions can
be treated essentially exactly using for example techniques such
as
bosonization\cite{solyom_revue_1d,emery_revue_1d,schulz_houches_revue,%
voit_bosonization_revue,senechal_bosonization_revue} so one only
needs good techniques to tackle the disorder. Quite interestingly
the physics of interacting and disordered one dimensional systems
is the one of disordered quantum solids. Higher dimensional
examples of such systems are the Wigner crystal, Charge Density
Waves and Bose glass. In these notes, we will explain the
techniques allowing to treat such systems, ranging from simple
physical arguments to a quite sophisticated variational approach.
In order to remain pedagogical and keep the algegra simple we
mostly discuss the technicalities on the simplest example of
spinless fermions.  We briefly discuss the specific physical
realizations and give references so that the reader can look in
more details the physical properties of these specific systems.

Before we embark with the physics, let us point out that these
notes results from the synthesis of various lectures. Some arbitrary choice of
material had to be made in order to keep some level of clarity.
Even if we have made some effort to cover various interesting topics,
these notes cannot pretend to be as exhaustive as a full
review. We thus apologize in advance to anybody whose pet problem
(or paper) is not covered in these few pages.

\section{Disordered interacting Fermions}

\subsection{Model}

Let us consider spinless fermions hopping on a lattice with
a kinetic energy $t$ and an interaction $V$
\begin{eqnarray} \label{eq:ham}
H &=& -t\sum_{i} (c_{i}^\dagger c_{i+1} + h.c.) \\
&+& \sum_{i> j} V_{i-j} (n_i-{\bar n})(n_j-{\bar n}) \nonumber
\end{eqnarray}
where $n_i$ ($\bar n$) is the local (average) electron density,
and the rest of the notation is standard ($t=1$ is the unit of
energy). For nearest neighbor interactions this is the well known
$t-V$ model. When using a Jordan-Wigner
transformation\cite{jordan_transformation} to express the
fermions in term of spins this latter model maps to an XXZ spin
chain. In addition to (\ref{eq:ham}) we want to submit the
fermions to a disorder. We concentrate here on site disorder.
Again most results/methods will carry over for randomness in the
hopping. The randomness is simply
\begin{equation}
H_{\rm int} = \sum_i \mu_i (n_i-{\bar n})
\end{equation}
where $\mu_i$ is a random variable of zero mean.

\subsection{Pure system}

The low energy properties of the pure system (\ref{eq:ham}) are
by now well understood. We will thus review the bosonization of
(\ref{eq:ham}) only very briefly to fix the notations and refer
the reader to the various reviews on the subject
\cite{solyom_revue_1d,emery_revue_1d,schulz_houches_revue,%
voit_bosonization_revue,senechal_bosonization_revue}. To get the
low energy physics, it is enough to focus on the excitations
aroung the Fermi points. In the continuum limit $i\to x$ this
amounts to express the fermion field in term of slow varying
(with respect to the lattice spacing) field of right (with
momentum close to $+k_F$) and left (with momentum close to
$-k_F$) movers
\begin{equation}
c^\dagger_i \to \psi^\dagger(x) = e^{-ik_F x} \psi_+^\dagger(x)
+  e^{ik_F x} \psi_-^\dagger(x)
\end{equation}
In term of these fields (\ref{eq:ham}) becomes
\begin{equation} \label{eq:cont}
H = - i\hbar v_F \int dx  (\psi_+^\dagger \partial_x\psi_+(x) -
\psi_-^\dagger \partial_x\psi_-(x)) + \int dx_1dx_2 V(x_1-x_2)
\rho(x_1) \rho(x_2)
\end{equation}
where the density reads
\begin{equation} \label{eq:dens}
\rho(x) = \psi^\dagger_+\psi_+ +\psi^\dagger_-\psi_- +
(e^{-i2k_Fx} \psi^\dagger_+\psi_- + {\rm h.c.})
\end{equation}
The remarkable feature in $d=1$ is that all the excitations of the
system can be reexpressed in term of the fluctuations of density.
If one introduces a field $\phi(x)$ describing the long wavelength part of
the density (\ref{eq:dens}) reads
\begin{equation}\label{eq:densbos}
\rho(x) = \rho_0 - \frac1\pi \nabla\phi(x) + \frac1{(2\pi\alpha)}
(e^{-i2k_F x + 2 \phi(x)} + {\rm h.c.})
\end{equation}
where $\alpha$ is a lattice spacing, $\rho_0=\overline{n}/\alpha$ and:
\begin{eqnarray}
\nabla\phi(r) &=& \sum_\pm \psi^\dagger_\pm(x)\psi_\pm(x) \\
\nabla\theta(r) &=& \sum_\pm \pm \psi^\dagger_\pm(x)\psi_\pm(x)
\end{eqnarray}
$\theta$ is a similar field but associated with the long wavelength part
of the current. $\phi$ and $\Pi=\frac1\pi\nabla\theta$ are canonically
conjugate. Both the Hamiltonian and the fermion operator can be
expressed in terms of these two fields
\begin{eqnarray}
H & = & \frac1{2\pi}\int dx \left[ uK (\nabla\theta(x))^2 + \frac{u}{K}
(\nabla\phi(x))^2 \right] \label{eq:hambos}\\
\psi_\pm(r) & = &
\frac1{\sqrt{2\pi\alpha}}e^{-i(\pm\phi(r)-\theta(r))}
\label{eq:single}
\end{eqnarray}
Using (\ref{eq:single}) in (\ref{eq:dens}) one easily recovers
(\ref{eq:densbos}).
For the free Hamiltonian (\ref{eq:ham}) (with $V=0$) one has in (\ref{eq:hambos})
$u=v_F$ and $K=1$. What makes the boson representation so useful is
the fact that even in the presence of interactions the bosonized form
of (\ref{eq:ham}) remains (\ref{eq:hambos}) but with renormalized
(Luttinger Liquid) parameters $u$ and $K$.

\subsection{Disorder}

Similarly the disorder can be rewritten in the boson representation.
It is natural to separate the Fourier components with wavevectors close
to $q\sim0$ ($\eta$) and $q\sim \pm 2k_F$ ($\xi,\xi^*$).
The disorder becomes thus
\begin{equation} \label{eq:disfermions}
H_{\rm dis} = \eta(x)[\psi^\dagger_+\psi_+ +\psi^\dagger_-\psi_-]
+(\xi(x)\psi^\dagger_+\psi_- + {\rm h.c.})
\end{equation}
where $\eta$ and $\xi$ are now two independent random potentials.
$\eta$ is real but when the system is incommensurate ($2k_F \ne
\pi$) $\xi$ is complex. The potentials $\eta$ and $\xi$  correspond
 respectively to the
forward scattering and the backward scattering of the fermions on
impurities. (\ref{eq:disfermions}) reads in the boson
representation
\begin{equation}
H_{\rm dis} = -\eta(x)\frac1\pi\nabla\phi(x)
+\left[\frac{\xi(x)}{(2\pi\alpha)}e^{i2\phi(x)} + {\rm
h.c.}\right]
\end{equation}
For incommensurate systems the forward scattering can easily be
absorbed by a gauge transformation on the fermions. In the boson
language this amounts to shift $\phi \to \phi -\frac{2K}{u} \int^x dy
\eta(y)$. Since $\xi$ is complex and hence has a random phase this
shift does not affect the backscattering term (this will be even
more explicit on the replicated Hamiltonian (\ref{eq:repaction}) below). Thus
the forward scattering can be accounted for completely. Its main
effect is to lead to an exponential decay of the disorder averaged density
correlation functions. Since the current and the superconducting
correlation functions depends only on $\theta$ (or equivalently in
an action representation on $\partial_t \phi$) they are not
affected by the shift. Transport properties thus depends on the
backscattering alone, as is obvious on physical basis since
forward scattering cannot change the current. For commensurate
potentials the situation is mode complicated since $\xi$ is real
($e^{+i2k_F}=e^{-i2k_F}$) and the forward scattering now {\it
does} affect the backscattering term (in other words by combining
a forward scattering and a scattering on the lattice one can
generate a backscattering term). This has consequences that we
will examine in more details in section~\ref{sec:comm}. For the moment let
us focus on the incommensurate case and get rid of the forward
scattering.

\section{Tackling the disorder} \label{sec:fukulee}

\begin{figure}
 \centerline{\psfig{file=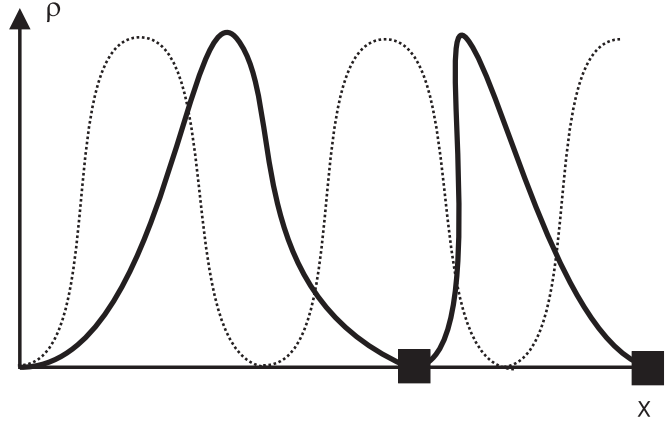,angle=0,width=\figwidth}}
  \caption{\label{fig:pincdw} Impurities (black square) will pin the density wave.
  The phase $\phi$ will have to change compared to the undistorted density (dotted
  line), which costs kinetic energy.}
\end{figure}

Disordered one dimensional particles are thus described by
\begin{equation}\label{eq:incaction}
S/\hbar = \int dx d\tau \left[ \frac1{2\pi K}\left[
\frac1{v}(\partial_\tau \phi)^2 + v (\partial_x \phi)^2
\right] + \frac{\xi(x)}{2\pi\alpha\hbar}  e^{i 2 \phi(x)} + {\rm h.c.}\right]
\end{equation}
where we have rewritten (\ref{eq:hambos}) and
(\ref{eq:disfermions}) as an action. We have reintroduced $\hbar$
and other pesky constants to show explicitly the various physical
limits. Note that although we are mainly concerned here with
fermions (\ref{eq:hambos}) describes in fact nearly every one
dimensional disordered problem ranging from dirty
bosons\cite{giamarchi_loc_lettre,giamarchi_loc,fisher_bosons_scaling}
to disordered spin chains\cite{doty_xxz}, since all this problems
have essentially the same boson representation. We will examine
these different systems in more details in
section~\ref{sec:othersystems}. (\ref{eq:incaction}) emphasizes
the physics of the problem. The electron system can be viewed as a
charge density wave\cite{gruner_revue_cdw,fukuyama_cdw_pinning}
since the density varies as (\ref{eq:densbos})
\begin{equation}
\rho(x,\tau) = \rho_0 \cos(2k_F x - 2 \phi(x,\tau))
\end{equation}
The elastic term in (\ref{eq:incaction}) wants the phase of the
density wave to be as constant as possible, and have a nice
sinusoidally modulated density. The disorder term on the contrary
wants to pin this charge density on the impurities by distorting
the phase, as shown on Figure~\ref{fig:pincdw}.
The problem of localization of interacting fermions is thus very
similar to the one of the pinning of classical charge density
waves\cite{fukuyama_pinning}.
The charge density wave is here intrinsic to the one-dimensional
interacting electron gas and not due to a coupling to phonons. The main
features are nevertheless similar, the main difference being the fact
that the effective mass  of the ``CDW'' is much smaller in the absence of the
electron-phonon coupling and hence the importance of the quantum
fluctuations is much higher. Various quantities of special interest
have a quite simple expression in terms of the bosons.
The conductivity is simply given by
\begin{equation} \label{conduc}
\sigma(\omega) = \frac{i}{\omega}\left[\langle \partial_\tau\phi
 \partial_\tau\phi\rangle_{i\omega_n,q=0}\right]_{i\omega_n\to\omega+i\delta}
\end{equation}
whereas the compressibility is given by a similar correlation function but
with differents limits
\begin{equation} \label{compress}
\kappa = \frac1{\pi^2}\langle \partial_x\phi\partial_x\phi\rangle_{i\omega_n=0,q}
\end{equation}
The correlations of the $2k_F$ part of the density and of the
superconducting order parameter $\psi^\dagger_+\psi^\dagger_-$
are respectively given by
\begin{eqnarray}
\chi_\rho(r) &=& \langle e^{i2\phi(r)} e^{-i2\phi(0)}\rangle \\
\chi_s(r) &=& \langle e^{i2\theta(r)} e^{-i2\theta(0)}\rangle
\end{eqnarray}
where $r=(x,\tau)$ and $\pi\Pi(x) = \nabla\theta(x)$.

Given the highly nonlinear nature of the coupling to disorder,
(\ref{eq:incaction}) is quite tough to solve.
On the other hand, if $\xi(x)$ was just a
constant we would have a simple sine-Gordon theory for which a great
deal is known. We will thus tackle the problem by increasingly
sophisticated methods.

\subsection{Chisel and Hammer} \label{scha}

In the absence of quantum fluctuations, $\phi$ would be a
classical field and we would have a good idea of what happens.
This is the way Fukuyama and Lee \cite{fukuyama_pinning} looked at
this problem. Such an approximation is of course very good for
``phononic'' charge density waves \cite{gruner_revue_cdw} since
the quantum term is $\Pi^2/(2 M)$ and thus very small. For
fermions this corresponds to the ``classical'' limit $\hbar \to
0$, $K\to 0$ keeping $\overline{K}=K/\hbar$ fixed, and thus to
very repulsive interactions. In that case we can ignore all
quantum fluctuations, and look for a {\it static} solution for
$\phi$. It is of course crucial for the existence of such solution
  that the disorder does not
depends on time. This solution $\phi_0(x)$ describes the static
distortion of the phase imposed by the random potential. In the
absence of kinetic energy $(\nabla\phi)^2$ , it would be easy to
``determine'' $\phi_0$. If we write the random field $\xi$ as an
amplitude $|\xi(x)|$ and a random phase $2\zeta$, the disorder
term writes
\begin{equation}
\int dx |\xi(x)|  e^{i2(\phi(x)-\zeta(x))} + {\rm c.c.}
\end{equation}
The optimum is thus for $\phi_0(x)$ to follow the random phase on
each point. For point like impurities located on random positions
$R_i$, $|\xi|$ would just be the strength of each impurity
potential and $\zeta = k_F R_i$. Thus $\phi_0(x) = \zeta(x)$ is
the generalisation to any type of disorder (and in particular to
the Gaussian disorder so dear to the theorist) of the physics
expressed in Figure~\ref{fig:pincdw}: get the density minimum at
each impurity. In presence of kinetic energy following the random
phase would cost too much kinetic energy. We do not know exactly
now how to determine the optimal $\phi_0(x)$ but we can do some
scaling arguments. Let us assume that $\phi$ remains constant for
a lengthscale $L_{\rm loc}$. On this lengthscale $\phi$ takes the
value that optimizes the disorder term, which now reads
\begin{equation} \label{eq:dissqr}
E_{\rm dis} = \left[\int_0^{\Loc} \xi(x)\right]e^{i 2 \phi} + {\rm c.c.}
\end{equation}
If we take for example a gaussian distribution for $\xi$
\begin{equation}
\overline{\xi(x)\xi^*(x')} = \D \delta(x-x')
\end{equation}
one gets because of the average of a complex random variable on a
 box of size $\Loc$
that the disorder contributes as
\begin{equation}
E_{\rm dis} = -\sqrt{\D \Loc} e^{i(2\phi_0 - 2\Xi)}
\end{equation}
where $\Xi$ is some phase. It clear that the optimum energy is reached
if $\phi_0$ adjusts to this (now unknown) phase. The global energy gain
now scales as $\sqrt{\Loc}$. Between two segments of size $\Loc$ the
phase has to distort to reach the next optimal value. The distortion
being of the order of $2\pi$ the cost in kinetic energy reads
\begin{equation}
E_{\rm kin} \propto \frac1{\Loc}
\end{equation}
minimizing the total cost shows that the length over which $\phi_0$
remains constant is given by
\begin{equation} \label{eq:fukulee}
\Loc \propto \left(\frac1{\D}\right)^{\frac13}
\end{equation}
This tells us that the system {\it does} pin on the impurities
and that below $\Loc$ the system looks very much like an
undistorted system. Since at the scale $\Loc$, $\phi_0$ varies
randomly the $2 k_F$ density density correlations will decay
exponentially with a characteristic size $\Loc$. It is thus very
tempting to associate $\Loc$ with the Anderson localization
length. Note that for the free fermion point $\Loc \propto 1/\D$
instead of (\ref{eq:fukulee}), so the above formula is clearly
missing a piece of physics when $K$ is not zero. Nevertheless from
this simple scaling argument we have obtained: (i) the fact that
classical CDW or very very repulsive fermions are pinned
(localized) by disorder; (ii) the localization length; (iii) the
fact that the ground state should contain a static distortion of
the phase due to the disorder. Unfortunately we have no other
information on $\phi_0$, which is certainly a drawback.

Even with our limited knowledge of the statics we can
nevertheless try to extract the
dynamics. Let us assume that all deformations of the phase which are
not contained in the static distortion are small and thus that we can
write
\begin{equation}
\phi(x,\tau) = \phi_0(x) + \delta\phi(x,\tau)
\end{equation}
with $\delta\phi(x,\tau)\ll \phi_0(x)$ in a very vague sense since we
deal with random variables. One can try to expand the random term in
power of $\delta\phi$
\begin{eqnarray} \label{fluctu}
S_{\rm dis} &=& \int d\tau dx |\xi(x)| \cos(2(\phi(x,\tau) -
\zeta(x)))\\
            &\simeq& -2 \int d\tau dx |\xi(x)| \cos(2(\phi_0(x)-\zeta(x)))
             (\delta\phi(x,\tau))^2 \nonumber
\end{eqnarray}
One can thus use in principle (\ref{fluctu}) to compute the
various physical quantities. Note that the conductivity
(\ref{conduc}) will {\it not} depend {\it directly} on the statics
solution $\phi_0$ since $\partial_t\phi_0=0$, so we can hope to
compute it. Of course the dependence of the fluctuations
$\delta\phi$ in $\phi_0$ is hidden in (\ref{fluctu}). If $\phi_0$
was following the random phase at every point, then the disorder
term would just lead to a mass term for the fluctations and the
optical conductivity would show a gap. In fact this is not true
at every point so (\ref{fluctu}) leads to a distribution of
masses for the fluctuations. Unfortunately the knowledge of
$\phi_0$ is too crude to compute the conductivity accurately and
depending on what exactly is $L_{\rm loc}$ one can find either a
gap, a non analytic behavior or a $\sigma(\omega) \sim \omega^2$
behavior at small frequencies\cite{fukuyama_pinning}. Based on
physical intuition Fukuyama and Lee opted for the
later\cite{fukuyama_pinning}, but the method shows its
limitations here and does not allow a reliable calculation of the
physical quantities. More precise calculations of $\phi_0$ and
the conductivity can be performed in the classical limit $K\to 0$
using a transfer matrix formalism \cite{feigelman_cdw_exact}.

One thing that can be obtained from the partial knowledge of the fluctuations
is the effect of quantum fluctuations\cite{suzumura_scha}.
Indeed when quantum fluctuations
are present the expansion (\ref{fluctu}) is not valid any more since the
cosine should be normal ordered before it can be expanded. This leads to
\begin{equation} \label{quantum}
\cos(2(\phi_0-\zeta + \delta\phi)) \simeq - 2\cos(2(\phi_0-\zeta))
e^{-2\langle \delta\phi^2\rangle} (\delta\phi)^2
\end{equation}
where the average $\langle\rangle$ has now to be computed self-consistently
using (\ref{quantum}). This leads to a modified localisation
length\cite{suzumura_scha}
of the form
\begin{equation} \label{schalength}
\Loc \propto \left(\frac1{\D}\right)^{\frac1{3-2K}}
\end{equation}
This expression for the localization length suggests that a delocalization
transition is induced by the quantum fluctuations and occurs at
$K=3/2$. In the fermion language this corresponds to extremely attractive
interactions.

\subsection{Starting from the metal: RG} \label{sec:rg}

The previous method starts directly from the localized phase. It
provides some limited information about this phase, but suffers
from serious limitations. An alternative approach is to start
from the pure Luttinger liquid and investigate the effects of
disorder perturbatively, and build a renormalization group
analysis. The RG provides us with the best possible description
of the delocalized phase and the critical properties of the
transition. It also gives a very accurate description of the
localized phase {\it up to} lengthscales of the order of the
localization length $\Loc$. Here again we describe the method for
simplicity on spinless fermions and discuss more complex systems
in section~\ref{sec:othersystems}.

To have a hint of the RG equations let us expand the disorder term
(\ref{eq:incaction}) to second order. This leads to
\begin{equation} \label{seconddis}
\int d\tau dx \int d\tau' dx' \xi(x)\xi^*(x') e^{2i (\phi(r)-\phi(r'))}
\end{equation}
It is easy to see that at the
tree level, (\ref{seconddis}) scales as $\D L^{3-2K}$. This leads to the
scaling of the disorder
\begin{equation} \label{rgdis}
\frac{\partial \D}{dl} = (3-2K) \D
\end{equation}
This traduces in fact the dressing of the scattering on the disorder by
the interactions and has been derived using either diagrams or RG
\cite{gorkov_pinning_parquet,mattis_backscattering,%
luther_conductivite_disorder,apel_spinless,apel_loc,apel_impurity_1d,%
giamarchi_loc_lettre,giamarchi_loc}.
In itself it seems to confirm the result of (\ref{schalength})
i.e. the existence of a transition. Note that the advantage of
the bosonization derivation is to allow to reach the non
perturbative point in interactions where such a metal-insulator
transition would take place.

In fact (\ref{rgdis}) would not allow in itself to really determine the
metal-insulator transition point. This can be seen by using the
RG to  compute the finite temperature (or finite
frequency) conductivity of the system
\cite{giamarchi_loc_lettre,giamarchi_loc}. The idea is simply to
renormalize until the cutoff is of the order of the thermal length $l_T
\sim u/T$ corresponding to $e^{l^*} \sim l_T/\alpha$.
At this length scale the disorder can be treated in the Born
approximation. As the conductivity is a physical quantity it is not
changed under renormalization and we have:
\begin{eqnarray} \label{renodens}
\sigma(n(0),D(0),0) = \sigma(n(l),D(l),l) = \sigma_0
\frac{n(l)D(0)}{n(0)D(l)} = \sigma_0 \frac{e^l D(0)}{D(l)}
\end{eqnarray}
where $\sigma(n(l),D(l),l)=\sigma(l)$ and $n(l)$ are respectively
the conductivity and the electronic density at the scale $l$.
$\sigma_0=e^2 v_F^2/2\pi\hbar \D$
is the conductivity in the Born approximation,
expressed with the initial parameters.  Using (\ref{rgdis})
one gets from (\ref{renodens})
\begin{equation} \label{simple}
\sigma(T) \sim \frac1{\D} T^{2-2K}
\end{equation}
This result is the direct consequence of the renormalization of
the scattering on impurities by interactions (\ref{rgdis}). One
immediately sees that (\ref{simple}) {\it alone} would lead to a
paradox since (\ref{rgdis})  gives a localized-delocalized
boundary at $K=3/2$ whereas (\ref{simple}) gives perfect
conductivity above $K=1$ (i.e. the noninteracting point). One
could also immediately see that if one introduces a new variable
such as
\begin{equation} \label{change}
\tilde{\D} = e^{-a l} \D
\end{equation}
the dimension of such a variable would be $(3 - a -2K)$, leaving
the location of the transition point  as determined from
(\ref{rgdis}) quite arbitrary. Although such a transformation
seems arbitrary, if one considers that the disorder stems from
impurities with a concentration $n_i$ and a strength $V$, the
limit of Gaussian disorder corresponds simply in taking $n_i\to \infty$
with $V \to 0$ keeping $\D = n_i V^2$ fixed. Thus the choice $a=1$
in (\ref{change}) simply corresponds to $\tilde{\D} = V^2$ i.e. writting an
RG equation for the impurity strength.

\begin{figure}
\centerline{\psfig{file=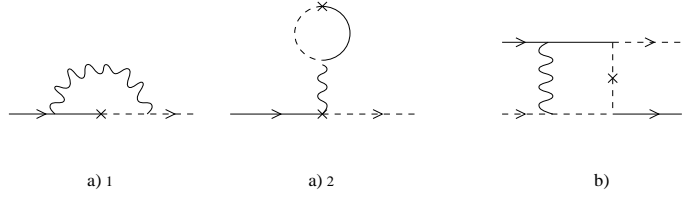,angle=-90,width=\figwidth}}
\caption{\label{fig:diagren}
Diagrams describing the renormalization of the disorder
by the interactions (a) and the renormalization of the interactions by
the disorder (b). Solid and dotted lines are fermions with $\pm k_F$,
the wiggly line is the interaction and the cross is the impurity
scattering.}
\end{figure}

\begin{figure}
\centerline{\psfig{file=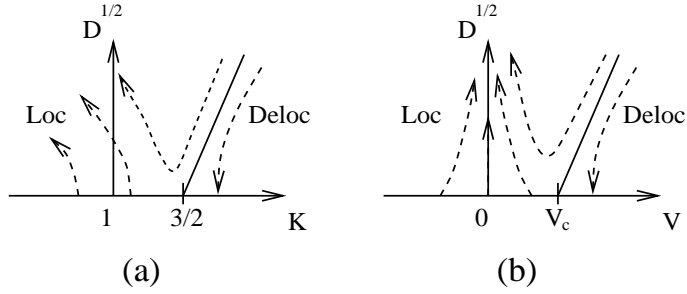,angle=0,width=\figwidth}}
\caption{\label{phasediag} Phase diagram and flow for spinless
fermions in presence of disorder. (a) is the flow in $\D$ and $K$
variables. (b) the flow in the $\D$ and interactions $g$.
Disorder kills inelastic interactions.}
\end{figure}

The answer to this simple paradox is of course that (\ref{rgdis})
should be complemented by another RG equation. In addition to
renormalizing $\D$ (\ref{seconddis}) generates as well quadratic
terms that renormalize the free part of the Hamiltonian, i.e. the
velocity $v$ and the Luttinger parameter $K$. Details can be
found in \cite{giamarchi_loc_lettre,giamarchi_loc}. The main
equation is the renormalization of the Luttinger parameter $K$
and reads
\begin{equation} \label{rgk}
\frac{\partial K}{dl} = -K^2 \D/2
\end{equation}
This equation describes the renormalization of the interactions by
the disorder. Both RG equations (\ref{rgdis}) and (\ref{rgk}) have
a diagramatic representation shown on figure~\ref{fig:diagren}.
Using the flow (\ref{rgdis}) and (\ref{rgk}) one can easily check that
two phases exist as shown on figure~\ref{phasediag}.
For large $K$ one in the delocalized phase where the disorder is
irrelevant and the system is a Luttinger liquid with renormalized
coefficients $u^*$ and $K^*$. All correlation functions decay as
power laws, and because $K^* > 3/2$ the system is dominated by
superconducting fluctuations. On the transition line the exponent
flows to the universal value $K^*=3/2$ Below this line $\D$ flows
to large values, indicating that the disorder is relevant. This
phase is the localized phase. This is obvious for physical
reasons but can also be guessed from the exact solution known for
the noninteracting line $K=1$ (and any $\D$) which belongs to
this phase. As can be seen from (\ref{rgdis}) and (\ref{rgk}) the
transition is Berezinskii-Kosterlitz-Thouless (BKT)
like\cite{kosterlitz_modele_xy,kosterlitz_renormalisation_xy} in
the $K$, $\sqrt{\D}$ variables.

In addition to the phase diagram itself a host of physical properties
can be extracted from the RG.
The simplest one is the localization length. One can use that for
$\D(l) \sim 1$
the localization length is of the order of the (renormalized) lattice spacing
$\alpha e^l$. The full determination needs an integration of both
(\ref{rgdis}) and (\ref{rgk}). Close to the transition the divergence
of the localization length  is BKT like (setting $K= 3/2 + \eta$)
\begin{equation}\label{eq:lloc-bkt}
L_{\rm loc} \sim e^{2\pi/\sqrt{9\D-\eta^2}}
\end{equation}
Deep in the localized regime, and for weak disorder, a good approximation
is to neglect the renormalization of $K$ in (\ref{rgdis}). A trivial
integration of (\ref{rgdis}) then gives back (\ref{schalength}).
This we see that the SCHA calculation corresponds in fact, both for the
phase diagram and for the localization length to the limit of
infinitesimal disorder.

Out of the RG one can also extract, using (\ref{renodens}) the behavior of
the temperature or frequency dependence of the conductivity. In the localized
phase this can only be used up to the energy scale corresponding to the
localization length i.e. $E_{\rm pin} = k_B T_{\rm pin}=\hbar
\omega_{\rm pin},\omega_{\rm pin} = v/L_{\rm loc}$.
Below this lengthscale another method than the weak coupling RG should be
used. We will come back to that point in section~\ref{variat}.
Here again, although the full flow should be taken into account one
can get an approximate formula by ignoring the renormalization of $K$,
which leads to  (\ref{simple}).
For $K < 3/2$ (including the noninteracting point) any small but {\bf
finite} disorder {\bf grows}, renormalizing the exponents
and ultimately leading to a decrease of the conductivity,
even if one started initially from $K > 1$. A very crude way of taking into
account both equations (\ref{rgdis}) and (\ref{rgk}) would be to say that one
can still use (\ref{simple}) but with scale dependent exponents (see
\cite{giamarchi_loc_lettre,giamarchi_loc})
\begin{equation} \label{complique}
\sigma(T) \sim T^{2-2K(T)}
\end{equation}
This renormalization of exponents and the faster decay
of conductivity is in fact the signature of Anderson localization.
The equivalent frequency dependence is
\begin{equation}
\sigma(\omega) \propto \omega^{2K-4}
\end{equation}
Here again this formula break down below the scale $\omega_{\rm pin}
\sim v/L_{\rm loc}$ which is the pinning frequency.
Similarly below $L_{\rm loc}$ correlation functions can
again be computed using the RG, but
of course the asymptotic behavior cannot be obtained.

\section{Other systems and RG} \label{sec:othersystems}

Despite its limitations to physics above $E_{\rm pin}$ the RG is an
extremely efficient method given its simplicity. It allows in addition
a perfect description of the delocalized phase and of the critical
behavior, something unatainable through the methods of section~\ref{scha}
and allows for interesting extensions.

First let us note that the equations (\ref{rgdis}) and (\ref{rgk}) also
describe the case of a single
impurity\cite{kane_qwires_tunnel_lettre,kane_qwires_tunnel}.  Indeed in that case one can
go back to the definition $\D = n_i V^2$ and take the limit $n_i \to 0$,
i.e. $\D\to 0$. (\ref{rgk}) shows that in that case $K$ cannot be renormalized
since a single impurity cannot change the thermodynamic behavior. Only
(\ref{rgdis}) remains, leading directly to temperature dependence of
the form (\ref{simple}), and a localized-delocalized transition at $K=1$.
More details on such a relation between the two problems and on the remaining
open question can be found in \cite{giamarchi_moriond}.

Quite remarkably the set (\ref{rgdis}-\ref{rgk}) seems wrong.
Indeed $K$ naively depends on the (inelastic) interations.
Perturbatively, for the pure systems $K=1-V/(2\pi v_F)$. If one
start for $K=1$, i.e. for the non interacting system, it would
thus seem from (\ref{rgk}) that the {\it elastic} scattering on
the impurities can generate {\it inelastic} fermion-fermion
interactions. The solution of this paradox is hidden in the
precise way the RG procedure is build. In order to have the
elastic nature of the scattering on impurities, the time
integrations in (\ref{seconddis}) should be dome independently for
$\tau$ and $\tau'$. When one performs the RG one introduces a
cutoff and imposes $|\tau-\tau'|>\alpha$. Thus a part is left out
of (\ref{seconddis}) which is
\begin{eqnarray}
\D\int dx \int_{|\tau-\tau'|<\alpha} d\tau d\tau'
\rho(x,\tau)\rho(x,\tau') \nonumber \\
\simeq 2\D\alpha \int dx \int d\tau \rho(x,\tau)\rho(x,\tau)
\end{eqnarray}
which is exactly an inelastic interaction term. Thus in fact $K$
contains not only the original inelastic interactions $V$ but
also a small correction coming from the disorder itself. In order
to determine the flow for $V$ it is thus necessary to take this
small correction into account
\cite{giamarchi_loc_lettre,giamarchi_loc} which gives the flow of
Figure~\ref{phasediag}-b. One thus sees that the elastic case
$V=0$ indeed remains elastic and also that for spinless fermions,
the perturbative flow seems to indicate that the inelastic
interactions are reduced by the disorder. This is compatible with
the physical image that one would get at strong disorder:
fermions localize individually and since the overlap of
wavefunctions is exponentially small, so is the effect of
interactions. One could thus naively expect that below $L_{\rm
loc}$ the effect of interactions are strong but disappear above
$L_{\rm loc}$. As we will see in section~\ref{variat} the
variational approach confirms this image. For fermions with spins
(to be discussed below) we expect only the interaction in the
charge sector to vanish. Interactions in the spin sector do
remain and lead to random exchange.

Of course many more physical systems can be studied by this method.
This is the case for spin chains, than can directly be mapped
onto spinless fermions\cite{jordan_transformation}.
 A spin chain under a random magnetic field along $z$
\cite{doty_xxz} is directly the problem that we solved in the
previous section, with $K$ being the anisotropy ($K=1$ for an XY
chain and $K=1/2$ for an Heisenberg one). However although  for
fermion problems it is unlikely that one is at a commensurate
filling this is the natural situation for a spin chain (since the
magnetization is zero in the absence of external field the
filling of the equivalent spinless fermion system is $n=1/2$). A
spin chain with random exchange is thus like a {\it commensurate}
fermionic system with random hopping. We will come back to this
peculiar case in section~\ref{sec:comm}. Quite remarkably, in one
dimension, bosons\cite{giamarchi_loc_lettre,giamarchi_loc} lead
to physics similar to fermions. This is due to the fact that in
one dimension statistics cannot be separated from interactions.
Interacting bosons can thus be represented by a bosonization
representation quite similar to the one of
fermions\cite{haldane_bosons}. The density can be written as
\begin{equation}\label{eq:density-bosonized-general}
\rho(r) = \rho_0 - \rho_0\partial_x\phi + \sum_{n} e^{i n
(2\pi\rho_0 x - 2 \phi(r))}
\end{equation}
very similar to the fermionic form.
while the single particle operator is now
\begin{equation}
\psi(r) \sim \rho_0^{1/2} e^{i \theta}
\end{equation}
(note the difference with the fermionic operator). The
hamiltonian is still described by the quadractic form
(\ref{eq:incaction}). Now $K=\infty$ for noninteracting bosons and
$K=1$ for hard core ones. There is thus a
transition\cite{giamarchi_loc_lettre,giamarchi_loc} between a
superfluid state (for $K>3/2$) to a Bose glass state where the
bosons are localized by the random potential at $K=3/2$.

In a similar way one can of course treat fermions with spins. Equations are
more complicated since they involve the charge and spin sectors, and we
will not discuss the full physics here but refer the reader
to \cite{giamarchi_loc}. Let us just draw attention to one interesting
consequence of the renormalization equation of the disorder for the
problem with spins, which reads
\begin{equation} \label{disorderspin}
\frac{d \D}{dl} = (3 - K_\rho - K_\sigma - g_{1\perp}) \D
\end{equation}
where $g_{1\perp}$ is the backscattering interaction between opposite
spins. For a Hubbard type interaction $g_{1\perp} = U$.
For spin isotropic systems $K_\sigma = 1 + g_{1\perp}/(2\pi v_F)$
and $g_{1\perp}$ is marginal with a flow
\begin{equation}
\frac{d g_{1\perp}}{dl} = - g_{1\perp}^2
\end{equation}
For more general spin couplings either $g_{1\perp} \to 0$ and
$K_\sigma \to K_\sigma^*$, or $g_{1\perp}$ is relevant and a spin
gap opens. The physics of the localization transition depends
thus on the sign of the interactions with special physical
consequences. The transition point move from $K_\rho = 2$ (for
infinitesimal disorder) for repulsive interactions to $K_\rho=3$
for attractive ones. For a Hubbard type interaction for which
$1/2 < K_\rho < 2$ the delocalization point can never be reached
and the system is localized regardless of the strength of
interaction and disorder. Another consequence is that in the
presence of spin degrees of freedom the divergence of the
localization length at the transition is not BKT like any more.
One could naively think that in the repulsive case the
$g_{1\perp}$ term could be omitted and that the renormalization
of the disorder could be written $(3-K_\rho - K_\sigma^*)$. For
spin isotropic interactions this misses an important part of the
physics.  Indeed, for a Hubbard interaction $K_\rho = 1 - U/(2\pi
v_F)$. Substituting in (\ref{disorderspin}) leads (for the
initial steps of the flow)
\begin{equation} \label{right}
\frac{d \D}{dl} = (1 - \frac{U}{\pi v_F}) \D
\end{equation}
whereas the incorrect substitution at the fixed point would lead
to $(1 + \frac{U}{\pi v_F})$, leading to quite different physics.
(\ref{right}) implies that for Hubbard type interactions
repulsive interactions make the system {\it less} localized
\cite{giamarchi_persistent_1D} than for attractive interactions,
i.e. that the $L_{\rm loc}^{U>0} > L_{\rm loc}^{U<0}$. Similar
effects exist for the charge stiffness and the persistent
currents, i.e. for a system with spins the persistent currents
are in fact enhanced by repulsive interactions. This
counter-intuitive statement can be explained physically:
interactions have two effects: (i) they tend to reinforce, when
attractive the superconducting fluctuations in the system. This
screens disorder and makes it less effective. This is the only
effect occurring for spinless fermions. (ii) when spin degrees of
freedom exists, repulsive interactions also tend to make the
density more uniform by spreading the charge. This makes it more
difficult to couple to disorder. These two effect compete. This
has several consequences, in particular for mesoscopic systems.
Of course, for fermions, true delocalization can only be achieved
with attractive interactions reaching at least the nearest
neighbor.

Many other systems have been treated by such method. there are
one-dimensional systems with long-range $1/r$ interactions, that
lead to a pinned Wigner crystal \cite{maurey_qwire}, doped spin 1
chains \cite{kawakami_dopeds=1}, fermionic
\cite{orignac_2chain_long} and bosonic
\cite{orignac_2chain_bosonic} ladders, spin 1 chains in a
magnetic disorder \cite{brunel_random_s=1}, spin ladders
\cite{orignac_2spinchains}. Since we want to focus here on the
methods we refer the reader to the above references for a
detailed discussion of the physics of such systems.

\section{A zest of numerics}\label{sec:numerics}

Although we are mainly concerned about analytical method in these
notes, let us mention some numerical results and methods that
have been used in connection with the RG predictions. Although
numerical studies have become very powerful in one dimension for
pure systems the presence of disorder complicates matters. Three
main methods have been used.

Exact diagonalizations, have been used to study both the phase
diagram and the charge stiffness of both spinless fermions
\cite{runge_xxz,bouzerar_spinless_currents} (or equivalently XXZ
spin chains) and fermions with spins
\cite{bouzerar_rg_current,romer_persistent_currents} with short
range or long range interactions
\cite{berkovits_coulomb_current}. Using the finite size scaling of
the spin stiffness $\rho_s=\frac{1}{L} \frac{\partial^2
E(\theta)}{\partial \varphi^2}=f\left(\frac{L}{L_{\rm
loc}}\right)$, where $E(\theta)$ is the ground state energy of
the disordered XXZ chain with boundary conditions $S^+_L=e^{i
\varphi}S^+_1$, the localization length $L_{\rm loc}$ can be
obtained \cite{runge_xxz}, and is in good agreement with the RG
results of section~\ref{sec:rg}. The behavior of the correlation
length close to the transition point appeared consistent with the
predicted BKT-like behavior. The results also suggested that a
finite disorder was needed to disorder the ground state for
$K>3/2$. A similar study with systems sizes of up to $L=18$ sites
was also made for XXZ spin chains with a random exchange in
\cite{haas_disorder_xxz}. Analysis of persistent currents was
also in agrement with the RG prediction of
section~\ref{sec:othersystems}. Unfortunately the exact
Diagonalization approach of the last section is limited to zero
temperature and small system size.

In order to consider bigger system sizes, one can use Quantum
Monte Carlo methods. In \cite{scalettar_bosons}, such a study was
performed for disordered bosons. The superfluid density was
obtained as a function of interaction for a given disorder
strength. It was shown that for not too repulsive equations, there
was a phase with a finite superfluid density. For more repulsive
interactions, a phase with finite compressiblity by zero
superfluid density was obtained, in agreement with the Bose Glass
theory of section~\ref{sec:othersystems}.

The most promising recent method is the Density Matrix
Renormalization Group. It been introduced in the recent years as
a method specially designed to calculate the ground state of
correlated one-dimensional systems\cite{white_dmrg}. This method
has been also applied to the problem of the XXZ chain in a random
magnetic field parallel to the z axis by Schmitteckert et al.
\cite{schmitteckert_dmrg}. The authors of
\cite{schmitteckert_dmrg} have been able to consider system size
of up to $L=60$ sites, and average over several hundred
realizations of the disorder. Localization and phase diagram were
also in good agreeement with the RG predictions.

Clearly, various numerical checks confirm the predictions of the
RG. Unfortunately so far only the phase diagram, stiffness and
localization length have been computed. This is clearly related to
the complexity of the problem at hand. What would be extremely
useful would be informations on quantities deep in the localized
phase such as the single particle Green's function, the ac or dc
conductivity. Analysis of such quantities would nicely complement
the RG analytical study and allow for comparison with other
analytical techniques more suited for the localized phase such as
the variational method we analyze in the next section.

\section{Variational Method} \label{variat}

Let us now study this problem using a completely different and at first
sight more formal method. As usual it is very convenient to get rid of
the disorder from the start. Given the non quadratic nature of
(\ref{eq:incaction}) supersymmetric methods are unapplicable and we
have to turn to replicas. The idea of the replica method in itself is quite
simple. If we want to compute an observable $O$ we have to do
both average over disorder and thermodynamic average
\begin{equation} \label{eq:repbas}
\overline{\langle O \rangle} = \int \D V p(V) \langle O \rangle_V
= \int D V p(V) \frac{\int \D\phi O[\phi] e^{-S_V[\phi]}}{\int D\phi
e^{-S_V[\phi]}}
\end{equation}
The action is usually linear in disorder and for Gaussian disorder the
distribution of random potential is $p(V) \propto e^{-\int dx V(x)^2}$,
so the average would be quite trivial without the denominator in
(\ref{eq:repbas}). The idea is thus to introduce $n$ fields and to
compute
\begin{eqnarray} \label{eq:replic}
\int D\phi_1 D\phi_2\ldots D\phi_n O[\phi_1] e^{-\sum_{i=1}^n S_V[\phi_i]}
= \nonumber \\
 \int  D\phi O[\phi] e^{-S_V[\phi]}\left[\int
D\phi e^{-S_V[\phi]}\right]^{n-1}
\end{eqnarray}
which is exactly the quantity we want to average over disorder in (\ref{eq:repbas})
if one takes the formal limit $n\to 0$. Since (\ref{eq:replic}) has no
denominator averaging over disorder is trivial. Of course there is a
price to pay: before the averaging the replicas are all independent
fields but the averaging introduces an interaction between them. We
have thus traded a theory depending on a random variable $V$ but a
single field for a theory without disorder but with $n$ coupled fields.
Usually this is still a situation we are better equipped to solve
because of the large number of field theoretic method dealing with
``normal'' (i.e. transitionally invariant actions). For the particular
case (\ref{eq:incaction}) the replicated action is
\begin{eqnarray}\label{eq:repaction}
S/\hbar &=& \int dx d\tau  \frac1{2\pi K}\sum_a\left[
\frac1{v}(\partial_\tau \phi_a)^2 + v (\partial_x \phi_a)^2
\right] - \\
& & \frac{\D}{(2\pi\alpha)^2\hbar} \sum_{ab} \int dxd\tau d\tau'
\cos(\phi_a(x,\tau)-\phi_b(x,\tau'))
\end{eqnarray}
where $a=1,\ldots,n$ is the replica index. Disorder averaging has
coupled the replicas via the cosine term. Because the disorder is time
independent this coupling contains two fields that can be at arbitrary
time and is thus highly non local. For fermions one usually prefers to
go to frequency space, where this imply conservation of the frequency
for each replica index, but this would not simplify things here because
of the cosine.

This is up to now a totally formal procedure and nothing has been
accomplished. (\ref{eq:repaction}) is totally equivalent to
(\ref{eq:incaction}) and the difficulty is of course to solve it.
Based on the RG equation (\ref{rgdis}) one could think naively
that since the localized phase corresponds to $\D \to \infty$ it
would be safe to expand the cosine term in (\ref{eq:repaction}).
Unfortunately it is easy to check that fails seriously {\it when
$n\to 0$ is taken} (it of course works perfectly for a finite
number of field $n \geq 2$). In order to circumvent this problem
let us try to improve over this simple minded expansion of the
cosine. Let us try a variational ansatz. We introduce a trial
action $S_0$
\begin{equation} \label{trialac}
S_0/\hbar =\frac1{2\beta L \hbar} \sum_{q,\omega_n}
\sum_{ab}\phi_a(q,\omega_n)G^{-1}_{ab}(q,\omega_n)\phi_b(-q,-\omega_n)
\end{equation}
where the propagators $G^{-1}$ are our variational ``parameters''.
As usual $\frac1L\sum_q \to \int\frac{dq}{(2\pi)}$.
If we introduce
\begin{equation}
Z = \int \D\phi e^{-S/\hbar}
\end{equation}
We then have the variational theorem for the free energy $F=-\hbar\log(Z)$
\begin{equation} \label{eq:trial}
F \leq F_{\rm tr} = F_0 + \langle S - S_0 \rangle_{S_0}
\end{equation}
Since $S_0$ is quadratic, (\ref{eq:trial}) can be in general computed
quite explicitely as a function of the (unknown) propagators $G$.
The ``best'' quadratic action $S_0$ is thus the one that satisfies the
saddle point equations
\begin{equation} \label{eq:saddle}
\frac{\partial F_{\rm tr}}{\partial G_{ab}(q,\omega_n)} = 0
\end{equation}
which gives a set of integral equations allowing to determine the
unknown functions $G$.

The observables are simply defined by quantities diagonal in
replica indices as can be seen from (\ref{eq:replic}). For some
quantities such as the compressibility it is necessary to be more
careful since one has to substract the average, which is usually
zero in a pure system or after averaging over disorder but non
zero for a {\it specific} realization of the disorder. Let us
introduce the various propagators (time ordering in $\tau$ is
always implied):
\begin{eqnarray}
B_{ab}(x,\tau) &=& \langle[\phi_a(x,\tau)-\phi_b(0,0)]^2\rangle
=  \nonumber \\
& & (G_{aa}(0,0) + G_{bb}(0,0) - 2 G_{ab}(x,\tau)) \label{lesb}\\
G_{ab}(q_x,\omega_n) &=& \langle
\phi_a(q_x,\omega_n)\phi_b(-q_x,-\omega_n) \rangle \label{lesg}
\end{eqnarray}
The compressibility is given by
\begin{eqnarray} \label{eq:compdis}
\chi(q,\omega_n) &=& \frac {1}{\hbar}\int dx \int_0^{\beta \hbar}
d\tau e^{-\imath (qx -\omega_n \tau)} \times \nonumber \\
& & \times \overline{\langle T_{\tau} (n(x,\tau)-\langle n(x,\tau)
\rangle)(n(0,0)-\langle n(0,0) \rangle) \rangle}
\end{eqnarray}
which leads to the average static compressibility $\chi_s=\lim_{q
\to 0}(\lim_{\omega \to 0} \chi(q,\omega))$ (see
(\ref{compress})). When expressed in terms of the replicated
bosonized operators (\ref{eq:compdis}) gives
\begin{equation}\label{replica_compressiblility}
\chi_s=\lim_{q\to 0} \lim_{\omega \to 0} q^2 G_c(q,\omega)
\end{equation}
where we introduced an important propagator: the connected one
defined as $G_c^{-1}(q) = \sum_b G_{ab}^{-1}(q)$.

Without the replicas this method is nothing but the well known
Self Consistent Harmonic Approximation (SCHA), which is known to
work very well for sine-Gordon type Hamiltonians. Such a method
gives in particular correctly the two phases (massless and
massive). Extension of this method to disordered systems was done
in the context of classical elastic systems such as interfaces
\cite{mezard_variational_replica}. In quantum problems another
level of complexity occurs because of the aforementioned non
locality of the interaction in time. But before going to these
problems, specific to the quantum systems, let us illustrate the
aspects of this variational method when applied to disordered
systems, on a technically simpler example (for which this method
was extremely fruitful \cite{giamarchi_book_young}): the case of
classical periodic systems.

\subsection{A classical example} \label{classical}

Let us take the action (\ref{eq:repaction}) but with only a {\it
single} time integral for the disorder term. Such action would be
the result of the average on a disorder both dependent on space
{\it and} time. Of course such a disorder would be quite
unrealistic for quantum problems. However
(\ref{eq:repactionclassical})  would be a perfectly natural
Hamiltonian for a {\it classical} problem where $z = v\tau$ is
now just one of the spatial dimensions
\cite{giamarchi_vortex_long,giamarchi_book_young}. To make the
analogy more transparent let us use $z=v\tau$, and replace the
integral over $x$ by an integral in $d-1$ dimensions. If denote
by $r$ the $d$-dimensional space variable $r=(x,z)$ the starting
action is
\begin{eqnarray}\label{eq:repactionclassical}
S/\hbar &=& \int d^dr  \frac1{2\pi K}\sum_a
(\partial_r \phi_a)^2  \\
&-& \frac{\D}{(2\pi\alpha)^2\hbar v} \sum_{ab} \int d^dr
\cos(2(\phi_a(r)-\phi_b(r))) \nonumber
\end{eqnarray}
One can see that (\ref{eq:repactionclassical}) is exactly
the Hamiltonian describing an elastic system such as a vortex lattice
or a classical CDW in the presence of point like defects in $d$ dimensions.
$\hbar$ plays the role of the temperature for the classical system,
the elastic constant $c$ is given by $c= 1/(\pi \overline{K})$ with
$\overline{K}=K/\hbar$ and
$\rho_0^2 \Delta/2 = \D\hbar/(2\pi\alpha)^2$ would be the
correlator of the classical disorder \cite{giamarchi_vortex_long}.

If we call $q=(q_x,\omega)$ the $d$-dimensional momentum, without
loss of generality, the matrix $G^{-1}_{ab}(q)$ can be chosen of
the form $G^{-1}_{ab}= c q^2 \delta_{ab} - \sigma_{ab}$. We
obtain by minimization of the variational free energy the saddle
point equations
\begin{equation} \label{bebe}
G_c^{-1}(q)  =  c q^2, \quad \sigma_{a \ne b} =
\frac{2\D}{(\pi\alpha)^2}  e^{-2 B_{ab}(r=0)}
\end{equation}
Using (\ref{lesb}-\ref{lesg}) one obtains
\begin{equation} \label{bebete}
B_{ab}(r) =  \hbar \int \frac{d^dq}{(2\pi)^d}
(G_{aa}(q) + G_{bb}(q) - 2 \cos(q r) G_{ab}(q))
\end{equation}
For this particular problem, the connected part is not affected by
disorder. This is the consequence of a hidden symmetry
(statistical tilt symmetry) of (\ref{eq:repactionclassical}),
whose disorder part is not affected by any {\it local} shift of
$\phi_a(r)$ such as $\phi_a(r) \to \phi_a(r) + f(r)$, where $f$
is an arbitrary function. Such a symmetry does not exist for the
time correlated disorder natural in a quantum problem, with
important physical consequences on which we will come back. The
only interesting equation here is thus the equation for the
off-diagonal part $\sigma_{a\ne b}$. Because of the locality of
the interaction term between replicas in
(\ref{eq:repactionclassical}) the self energy $\sigma_{ab}$ is
simply a matrix of constants.

Given the symmetry of the original action/Hamiltonian
(\ref{eq:repactionclassical}) by permutation of the replica indices, it
is very natural to look for a variational ansatz with the same
symmetry. This would mean that the $G^{-1}$ matrix
would have only (for each value of $q$) {\it two} independent values:
the diagonal one $G_{aa} = \tilde{G}$ and the off diagonal one $G_{a\ne
b}$. Such a matrix can easily be inverted for any $n$, and the analytic
continuation for $n\to 0$ gives
\begin{eqnarray} \label{analytic}
G_c &=& \tilde{G} - G_{a\ne b} = \frac1{G^{-1}_c} \\
G_{a\ne b} & = & - \frac{G^{-1}_{a\ne b}}{(G^{-1}_c)^2}
\end{eqnarray}
Using these inversion formulas is it easy to solve for
(\ref{bebe}). In $d > 2$, $B$ depends on $G_c$ only and thus
$\sigma_{a\ne b}$ is simply a constant proportional to disorder.
Given the gaussian nature of the trial action (\ref{trialac}) the
correlation functions such as the density-density can easily be
computed
\begin{equation} \label{eq:densityreplica}
\overline{\langle \rho(r) \rho(0) \rangle} =
\langle \rho_a(r) \rho_a(0) \rangle \propto e^{-2B_{aa}(r)}
\end{equation}
Using (\ref{analytic}) shows that $G_{aa}(q) \sim \D/q^4$ leading
to a growth
\begin{equation} \label{growthrs}
B_{aa}(r) \sim \D r^{4-d}
\end{equation}
Although this solution is perfectly well behaved, a stability
analysis of the replica symmetric saddle point shows that it is
unstable. This can be checked from the eigenvalue $\lambda$ of
the replicon mode
\cite{mezard_variational_replica,giamarchi_vortex_long}.
\begin{equation} \label{replicon}
\lambda=1 -
\frac{8\hbar \D}{(\pi\alpha)^2} e^{- 4\hbar\int \frac{d^dp}{(2\pi)^d}
G_c(p)} \int \frac{d^dq}{(2\pi)^d} G_c^2(q)
\end{equation}
A negative eigenvalue $\lambda$ indicates an instability of the
replica symmetric solution. We introduce a small regularizing
mass in $G_c$: $G_c(q)^{-1}=c q^2 + \mu^2$ and take the limit
$\mu \to 0$. It is easy to see from (\ref{replicon}) that for
$d<2$ the replica symmetric solution is always stable. In that
case disorder is in fact irrelevant, due to the strong quantum
(or thermal for the associated classical system) fluctuations.
For $d=2$ the condition becomes $\mu^{2(\kbar - 1)}<1$ for small
$\mu$. Thus there is a transition at $\kbar = 1$ between a
replica symmetric stable high temperature phase where disorder is
irrelevant and a low temperature (glassy) phase where the
symmetric saddle point is unstable. For the classical system this
is the well known Cardy-Ostlund transition
\cite{cardy_desordre_rg}, with very interesting physical aspects
of its own. This transition is the equivalent for {\it time
dependent} disorder of the localization transition studied in
section~\ref{sec:rg}. For time dependent disorder one time
integral drops in (\ref{seconddis}) and (\ref{rgdis}) would
become $(2-2K)\D$ giving the transition at $K=1$. More details can
be found in
\cite{giamarchi_vortex_long,ledoussal_rsb_prl,giamarchi_book_young}.
For $2<d<4$ the replica symmetric solution is {\it always
unstable}.

\begin{figure}
\centerline{\psfig{file=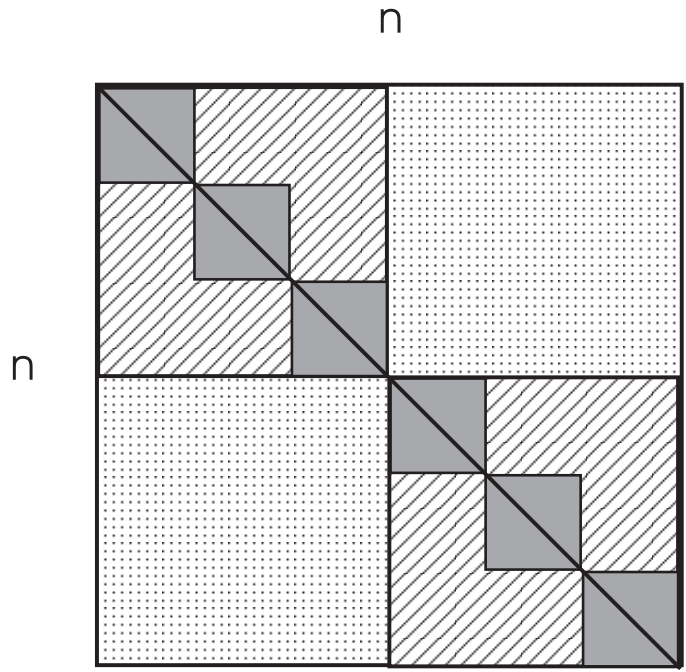,angle=0,width=\figwidth}}
\caption{Replica symmetry broken matrices (here for a 2-step
RSB). Each pattern correspond to a different value $\sigma_i$
(see text), and the diagonal one (black line) is
$\tilde{\sigma}$.}\label{fig:rsbpic}
\end{figure}

\begin{figure}
\centerline{\psfig{file=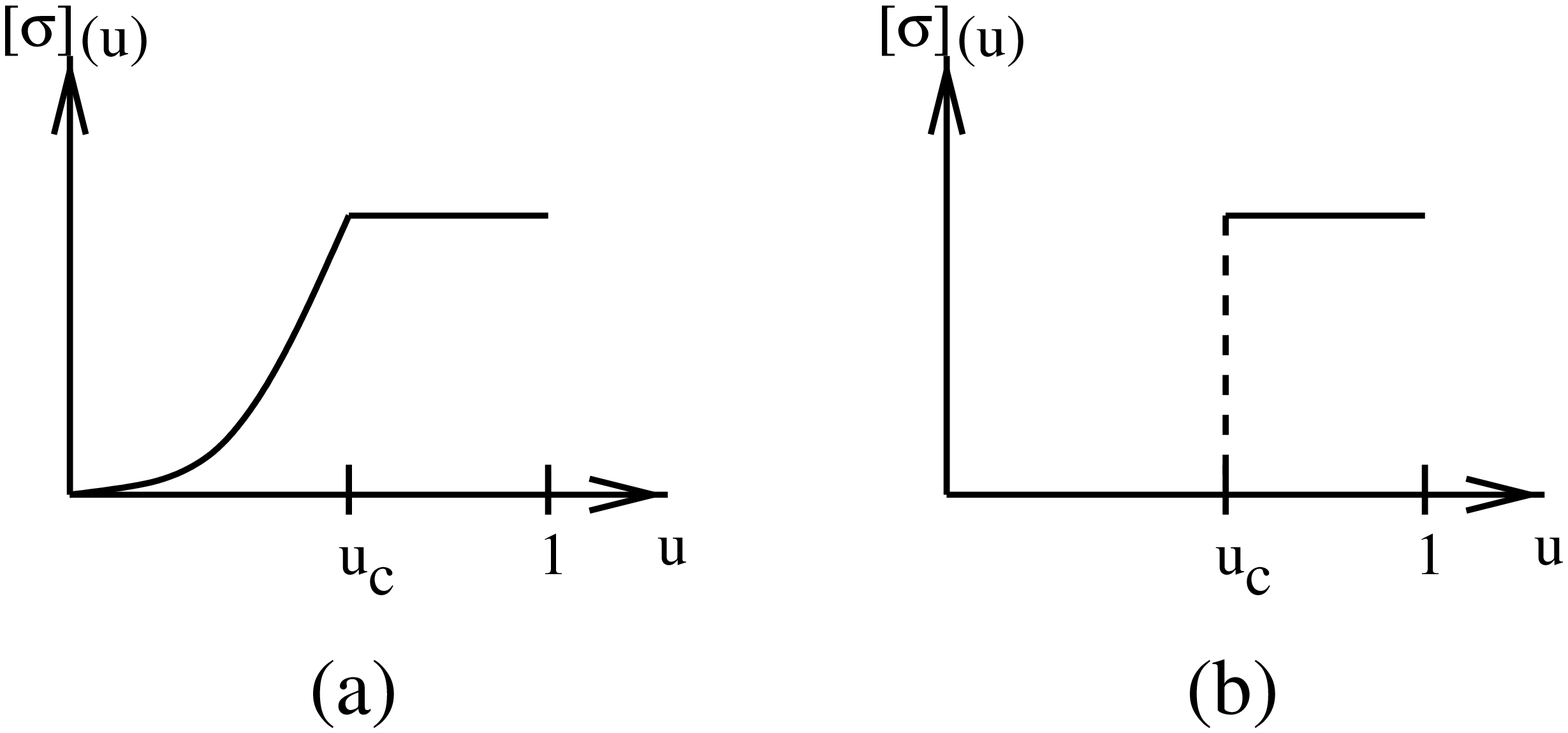,angle=0,width=\figwidth}}
\caption{\label{sigmapar}
Shape of the self energy $\sigma$ as a function of $u$. (a) is the
full RSB solution occuring for $2<d<4$. (b) a one step RSB ($d=2$).}
\end{figure}

One should thus look for another way of inverting the $0\times 0$
matrices than the replica symmetric one. Fortunately such a
scheme was invented in the context of spin glasses. Instead of
having a single value $\sigma$ for the off diagonal term, one
introduces a whole set of values. Let us briefly illustrate the
procedure here, refering the reader to \cite{mezard_book} for
details. Let us introduce a set of integers $m_0=n$, $m_1$, ...,
$m_{k+1}=1$ such that $m_i/m_{i+1}$ is an integer. One cuts the
matrix in blocks of size $m_1$ as illustrated on
Figure~\ref{fig:rsbpic}.
Elements outside the blocks have the value $\sigma_0$. The
procedure is then recursively applied for the inner blocks. At
the last step the value on the block of size $m_k$ is $\sigma_k$
and a diagonal value $\tilde{\sigma}$. Quite generally the matrix
is thus now parametrized by a diagonal element and a whole
function (in the limit $n\to 0$) $\sigma(u)$, where $u\in [0,1]$
(notice the range of variation of the $m_i$ when $n\to 0$). Such
matrices can also be inverted in the limit $n\to 0$), albeit with
more complicated inversion rules than for the RS solution. When
one has a continuous function (see Figure~\ref{sigmapar}) the RSB
is said to be continuous. Simpler cases are of course a constant
function (a single off-diagonal value) which is simply the RS
solution or a function continuous by steps. We have represented in
Figure~\ref{fig:rsbpic} (see also Figure~\ref{sigmapar}) the case
of a two-step RSB. We denote $\tilde{G}(q)=G_{aa}(q)$, similarly
$\tilde{B}(x) = B_{aa}(x)$, and parametrize $G_{ab}(q)$ by
$G(q,v)$ where $0<v<1$, and $B_{ab}(x)$ by $B(x,v)$. Physically,
$v$ parametrises pairs of low lying states, in the hierarchy of
states, $v=0$ corresponding to states further apart. The saddle
point equations become:
\begin{equation} \label{saddle}
\sigma(v)= \frac{2\D}{(\pi\alpha)^2} e^{- 2 B(0,v)}
\end{equation}
where
\begin{equation}
B(0,v) = 2\hbar \int \frac{d^dq}{(2\pi)^d} (\tilde{G}(q) -G(q,v))
\end{equation}
$B(0,v)$ corresponds physically to the mean squared phase
fluctuations at the same point in space ($r=0$) (for the
associated classical system this bould be mean squared relative
displacements of the same object) but in two replica states, or
more physically in  two different low lying metastable states.
The large distance behaviour of disorder-averaged correlators is
determined by the small $v$ behaviour of $B(0,v)$. We look for a
solution such that $\sigma(v)$ is constant for $v>v_c$, $v_c$
itself being a variational parameter, and has an arbitrary
functional form below $v_c$. This corresponds to full RSB (see
Figure~\ref{sigmapar}). The algebraic rules for inversion of
hierarchical matrices \cite{mezard_variational_replica} give:
\begin{equation} \label{inversion}
B(0,v)=B(0,v_c)+ \int_{v}^{v_c} dw \int \frac{d^dq}{(2 \pi)^d}
{2 \hbar \sigma'(w) \over {(G_c(q)^{-1} + [\sigma](w))}^2 }
\end{equation}
where $[\sigma](v)=u\sigma(v)-\int_{0}^{v} dw \sigma(w)$ and
\begin{equation}
B(0,v_c) = \int \frac{d^dq}{(2\pi)^d} \frac{2 \hbar}{G_c(q)^{-1} +
[\sigma](v_c)}
\end{equation}
This is a simple number.
Taking the derivative of (\ref{saddle})  with respect to $v$, using
$[\sigma]'(v)=v\sigma'(v)$, (\ref{inversion}), and
(\ref{saddle}) again one finds
\begin{equation} \label{relation}
1 = \sigma(v) \int \frac{d^dq}{(2 \pi)^d}
\frac{4\hbar}{(c q^2 + [\sigma](v))^2} \simeq
\sigma(v) \left(\frac{4\hbar  c_d}{c^{d/2}}\right)[\sigma(v)]^{(d-4)/2}
\end{equation}
Since the integral is ultraviolet convergent,
we have taken the short-distance momentum cutoff  to
infinity. $c_d$ is a simple number
\begin{equation}
c_d = \int \frac{d^d q}{(2 \pi)^d} \left(\frac1{q^2+1}\right)^2
= \frac{(2-d) \pi^{1 - d/2}}{2^{d+1} \sin(d\pi/2) \Gamma(d/2)}
\end{equation}
with $c_{d=3}=1/(8 \pi)$, $c_{d=2}=1/(4 \pi)$.
Derivating one more time
one gets for the effective self energy:
\begin{equation}\label{sigmeq}
[\sigma](v)=  (u/u_0)^{2 / \theta}
\end{equation}
where $\theta = (d-2)$ and $v_0 = 8 \hbar c_d c^{-d/2}/(4-d)$.
The shape of $[\sigma](u)$ is shown on Figure~\ref{sigmapar}.
The solution (\ref{sigmeq}), is a priori valid up to a breakpoint $u_c$,
above which $[\sigma]$ is constant, since $\sigma'(u) = 0$ is also a
solution of the variational equations. $u_c$ can also be extracted
from the saddle point equations and we refer the reader
to \cite{giamarchi_vortex_long} for details. The precise value of $u_c$ is
unimportant for our purpose but the existence of the two distinct
regimes in $[\sigma](u)$ has a simple physical interpretation that
we examine in section~\ref{aint}
Using (\ref{sigmeq})
one can now compute the correlation functions.
Larger distances correspond to less massive modes and is dominated by the
small $u$ behavior of (\ref{sigmeq}). One obtains
\begin{eqnarray}  \label{btildedef}
\overline{ \langle (\phi(r) - \phi(0))^2 \rangle } &=&
2 \hbar \int {d^dq \over (2\pi)^d}
(1 - \cos(qr) ) \tilde{G}(q) \\
\tilde{G}(q) &=& { 1\over{c q^2}} ( 1 + \int_{0}^{1} {dv \over v^2}
{ [\sigma](v) \over { c q^2 + [\sigma](v) } } ) \sim \frac{Z_d}{q^d}
\label{gtilde}
\end{eqnarray}
with $Z_d=(4-d)/(4\hbar S_d)$ and $1/S_d=2^{d-1}\pi^{d/2}\Gamma[d/2]$.
Thus for $2<d<4$ this leads to a logarithmic growth,
\begin{equation} \label{meansq}
\overline{ \langle (\phi(x) - \phi(0))^2 \rangle }=\frac{1}{2} A_d \log|x|
\end{equation}
with $A_d=4-d$, instead of the
power law growth (\ref{growthrs}) of the replica symmetric solution.
Note that the amplitude is independent of disorder.

\subsection{If it ain't broken ...} \label{aint}

It is thus necessary to break the replica symmetry to get the
correct asymptotic physics ($1/q^d$ propagator) instead of the
$1/q^4$ given by the replica symmetric solution. From
(\ref{gtilde}) one easily sees that at large enough $q$ (i.e. for
short distances) one recovers the replica symmetric solution.
Thus $u_c$ and $[\sigma(u_c)]$ (see figure~\ref{sigmapar}) define
a lengthscale $L \sim (c/[\sigma(u_c)])^{1/2}$ above which the RS
solution does not describe the physics. This lengthscale
corresponds to the pinning length similar to (\ref{eq:fukulee})
obtained by balancing the elastic energy with the disorder one.
Here
\begin{eqnarray}
E_{\rm el} &\sim& L^{d-2} \\
E_{\rm dis} &\sim& \D^{1/2} L^{d/2}
\end{eqnarray}
leading to the famous Larkin
\cite{larkin_70,larkin_ovchinnikov_pinning} length $L_{\rm loc}
\sim (1/\D)^{1/(4-d)}$ which corresponds to the distance for
which relative displacements are of the order of the lattice
spacing \cite{rcra} (or for which the phase here is of the order
$2\pi$) \cite{loflim_footnote}. Below this lengthscale the system
has a single equilibrium state. It can be described by simply
expanding in the displacements in (\ref{eq:incaction})
\begin{equation}
H_{\rm dis} = \int d^dx f(x) \phi(x)
\end{equation}
where $f$ is a random force, derivative of the random potential $V$.
It is easy to see that this model gives the $1/q^4$ propagator. This
breaks down when the displacements cannot be expanded, i.e. for
distances larger than $L_{\rm loc}$.
As anticipated by Larkin and
Ovchinikov\cite{larkin_ovchinnikov_pinning}, pinning occurs
that tends to keep a domain of size $L_{\rm loc}$ in place despite the
thermal fluctuations.

\begin{figure}
\centerline{\psfig{file=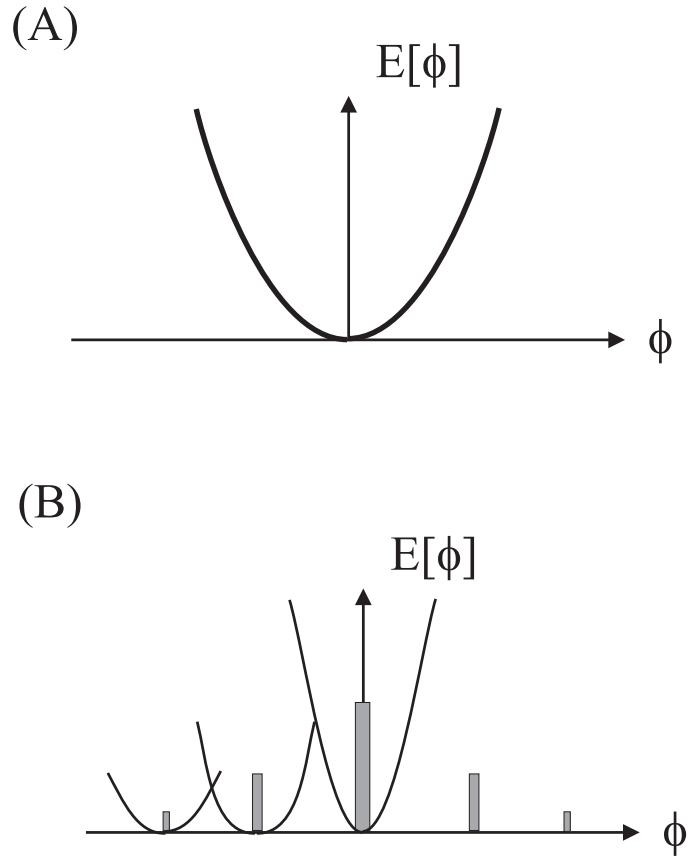,angle=0,width=\figwidth}}
\caption{\label{fig:rsvsrsb} (A) The RS solution describes
fluctuations around a single state, with a single spring constant.
(B) the RSB solution takes correctly into account the fact that
different regions of space correspond to different equilibrium
values of $\phi$, with a priori different spring constants. The
weight of a given configuration has been represented by the
height of the shaded region.}
\end{figure}

This help us to understand the physics of the RSB solution. To
illustrate it let us focus for simplicity to the case in $d=2$.
As can be seen from (\ref{sigmeq}) by letting $d\to 2$ the RSB
solution in that case is a one-step breaking as shown in
Figure~\ref{sigmapar}. There are some additional complications
but they are unimportant for our discussion here. A normal self
consistent approximation would approximate the energy by a simple
Gaussian centered in $\phi=0$. The only way to incorporate the
pinning is thus to put a mass term in the propagator
\begin{equation}
q^2 \to q^2 + L_{\rm loc}^{-2}
\end{equation}
As can be immediately inferred from the discussion of
section~\ref{scha}, this is a much too crude approximation. The
RSB solution is smarter and approximates effectively the
distribution of displacements by a hierarchical superposition of
Gaussians centered at different randomly located points in space.
A pictorial view of these two cases is shown on
Figure~\ref{fig:rsvsrsb}.
The double distribution over environment and thermal (or quantum)
fluctuations is approximated as follows
\cite{giamarchi_book_young}. In each environment there are
effective ``pinning centers'' corresponding to the low lying
metastable states (preferred configurations). Since all $q$ modes
are in effect decoupled within this approximation, for each $q$
mode a preferred configuration (a state) is $\phi_\alpha(q)$. They
are distributed according to:
\begin{eqnarray}
P(\{\phi_\alpha(q)\}) \sim \prod_{\alpha} e^{ -\frac{c}{2 T_g} q^2 |\phi_\alpha(q)|^2 }
\end{eqnarray}
Each is endowed with a free energy $f_\alpha$ distributed
according to an exponential distribution $P(f) \sim \exp(u_c f/T)$
(here $u_c=T/T_g$). Once these seed states are constructed, the
full thermal distribution of the $q$ mode $\phi_q$ is obtained by
letting it fluctuate thermally around one of the states:
\begin{equation}
P(\phi_q) \sim \sum_{\alpha} W_\alpha
e^{-\frac{c}{2 T} (q^2 + L_{\rm loc}^{-2}) |\phi_q - \phi_\alpha(q)|^2}
\end{equation}
where each state is weighted with probability
$W_\alpha=e^{-f_\alpha/T}/\sum_\beta e^{-f_\beta/T}$. One thus
recovers qualitatively the picture of Larkin Ovchinikov as the
solution of the problem with the replica variational method. The
Larkin length naturally appears as setting the (internal) size of
the elastically correlated domains. The full RSB case corresponds
to more levels in this hierarchy of Larkin domains (in some sense
there are clusters of domains of size larger than $L_{\rm loc}$)
and the way this hierarchy scales with distance reproduces the
exponents for displacements and energy fluctuations. We refer the
reader to \cite{giamarchi_book_young} for more details on the use
of the variational method in this context and for the physical
properties of such systems.

\subsection{Quantum problems}\label{sec:quantumvaria}

We now apply the same method to the quantum problem. Because of the
non locality in time of the interaction in (\ref{eq:repaction}) the solution
will have quite different properties.
We again use the trial action (\ref{trialac}),
with the parametrization
\begin{equation}\label{eq:greenvaria}
v G^{-1}_{ab}(q,\omega_n)=\frac {((vq)^2+\omega_n^2)}{\pi K}\delta_{ab} -
\sigma_{ab}(\omega_n)
\end{equation}
Although, as in section~\ref{classical} $\sigma_{ab}$ is still independent of $q$
because of the locality in space it is now dependent on $\omega_n$. This
would render the solution extremely complicated if it were not for a
remarquable property of quantum systems. Off diagonal replica terms
such as $\sigma_{a\ne b}$ {\it only} exist for the mode $\omega_n=0$
The general argument \cite{subir_replicas,giamarchi_columnar_variat} is that
in each realization of the random potential $V$, the disorder does not depend
on $\tau$. Therefore before averaging over disorder:
\begin{equation} \label{timeind}
G_{ab,V} = \langle \phi_a(x,\tau) \phi_b(0,0)\rangle =
           \langle \phi_a(x,\tau)\rangle \langle\phi_b(0,0)\rangle =
           \langle \phi_a(x,0)\rangle \langle \phi_b(0,0)\rangle
\end{equation}
It is important to note that such a property crucially depends on
the assumption that the hamiltonian is $\tau$-independent and
on the fact that equilibrium has being attained. This is the case
considered here.

The static mode $\omega_n=0$ thus plays a special role. This is
quite natural in a time independent disorder. Quite naively one
sees already that the properties of the variational solution will
thus be very similar to the ones of {\it point like} (i.e.
totally uncorrelated) disorder in $d$ spatial dimensions (here
$d=1$). This suggest strongly that the variational method will
pull out a static solution, reminiscent of the one introduced in
section~\ref{scha} and treat the fluctuations around this static
solution. We will come back to this point later. The solution can
be obtained quite in all dimensions and we refer the reader to
\cite{giamarchi_columnar_variat} for details. We specialize here
to the case $d=1$. In this case two type of solution exist, a
simple RS solution with $\sigma_{ab} = 0$. This solution is
stable for $K > 3/2$. It corresponds of course to the delocalized
regime where the cosine term in (\ref{eq:repaction}) is
irrelevant. The variational method  correctly reproduces the
(gaussian) delocalized regime, and the correct transition point,
but of course misses the renormalization of the Luttinger
parameters given by the RG. For $K < 3/2$ although an RS solution
still exists it is unstable and physically obviously incorrect
\cite{giamarchi_columnar_variat}. One should look for an RSB
solution. In that case the correct solution is a one-step RSB
solution (it would be full RSB for $d>2$) of the type shown in
Figure~\ref{sigmapar}.
\begin{eqnarray}\label{saddle-point-backward}
G_c^{-1}(q,\omega_n) & = & \frac \hbar {\pi K} (v
q^2+\frac{\omega_n^2} v) + \frac{2\D}{\hbar(\pi \alpha
)^2}\int_0^{\beta \hbar}d\tau
(1-\cos(\omega_n \tau)) \nonumber \\
& & \left[\exp(-2\hbar \tilde{B}(x=0,\tau))-\int_0^1 du \exp (-2
\hbar B(u))\right]
\end{eqnarray}
with
\begin{eqnarray}
\sigma(q,\omega_n,u)=\frac{2 \D v }{ (\pi \alpha
)^2}\beta \exp (-\hbar 2 B(u))\delta_{\omega_n,0}
\end{eqnarray}
These equations are still formidable to solve. A simple
parametrization is \begin{eqnarray} \label{eq:gcrsb}
vG_c^{-1}(q,\omega_n)=\frac 1 {\pi \overline{K}}((v
q)^2+\omega_n^2) + \Sigma_1(1-\delta_{n,0})
+I(\omega_n) \\
I(\omega_n)=\frac{2 \D v}{ (\pi \alpha )^2 \hbar
}\int_0^{\beta\hbar} \left[e^{-2\hbar
\tilde{B}(\tau)}-e^{-2\hbar  B(u>u_c)}\right](1-\cos(\omega_n
\tau)) d\tau \label{eq:iomegarsb} \\
\Sigma_1=u_c(\sigma(u>u_c)-\sigma(u<u_c))=[\sigma](u>u_c)\label{eq:Sigmarsb} \\
\sigma(u)=\frac{2 \D v}{ (\pi \alpha)^2}e^{-\hbar 2 B(u)}\beta
\delta_{n,0}\label{eq:sigmaursb}
\end{eqnarray}
The parameters $\Sigma_1$, the breakpoint $u_c$ and the
function $I(\omega_n)$ have to be determined self-consistently.
Let us examine first the general properties of the solution.

\begin{figure}
\centerline{\psfig{file=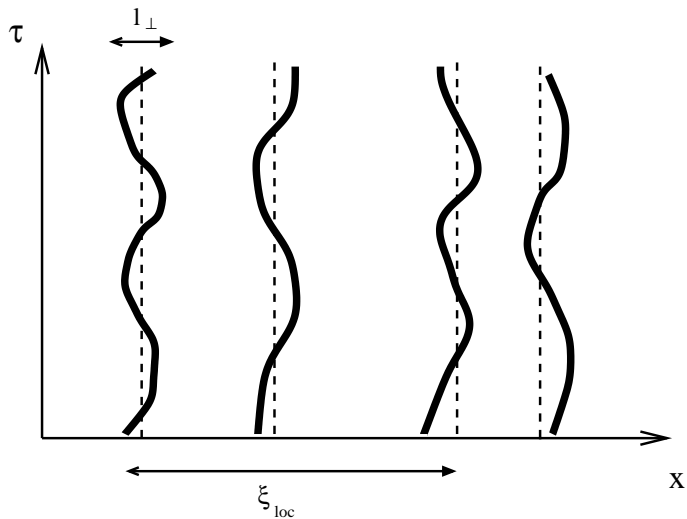,angle=0,width=\figwidth}}
\caption{\label{widthfluct}
Localization.}
\end{figure}

Since $I(\omega_n=0)=0$ it is easy to check from (\ref{compress}) that
the compressibility is unchanged by the disorder, since the ``mass''
term $\Sigma_1$ goes also away {\it at} $\omega_n=0$. The variational
method thus correctly reproduces that the compressibility of an
Anderson insulator is still finite, and practically unchanged (for
free electrons where we can compute it) from the value without
disorder. Correlation functions are also
easy to obtain. We just give here
the important physical point without the explicit derivation
\cite{giamarchi_columnar_variat}.
Because of the presence of $\Sigma_1$ in the propagator
(\ref{eq:gcrsb}) they will be massive. This leads to
\begin{equation}
B(x\to \infty, \tau=0) \to x/L_{\rm loc} \label{localspace}
\end{equation}
A full calculation of the correlations shows that
$\tilde{B}(\tau)$ grows until $\tau \sim L_{\rm loc}$ when
it saturates
\begin{equation}\label{localtime}
B(x=0,\tau\to\infty)   \to {\rm Cste} = (\rho_0^{-1} l_\perp)^2
\end{equation}
Since the variational action is gaussian one has for the
correlation of the $2k_F$ part of the density
\begin{equation} \label{expodecay}
\chi_\rho(x,\tau) = e^{-2B(x,\tau)}
\end{equation}
Two different physical effects are described by the above
correlation functions. In the absence of disorder
$\chi_\rho(\tau\to\infty)$ would go to zero as a power law, a
sign of the wandering of the particles. (\ref{localtime})
traduces the fact that the time independent potential localizes
the particles at a given point in space, instead of letting them
fluctuate (unboundedly) due to quantum fluctuations in the
absence of disorder. This leads to density correlation in time
going to a constant up to a Debye Waller like factor. The
behavior (\ref{localtime}) thus shows that a {\it static}
solution $\phi_0(x)$ exists. The variational approach thus
provides a good justification for the static solution $\phi_0$ on
which the method of section~\ref{sec:fukulee} is built. The
correlation length of this static solution is given by
(\ref{localspace}). Since this length controls through
(\ref{expodecay}) the  exponential decay of the spatial
correlations of density it is thus related to the standard
localization length. Note that here both $L_{\rm loc}$ and {\it
the full static solution} are determined by the variational
method. Simple dimensional analysis on (\ref{eq:gcrsb}) shows that
\begin{equation}
L_{\rm loc} \propto 1/\sqrt{\Sigma_1}.
\end{equation}
The solution of the variational
equations \cite{giamarchi_columnar_variat} leads back to the
expression (\ref{schalength}) for $L_{\rm loc}$.
Quite interestingly (\ref{localtime})
defines a length if one writes the constant in units of the fermion
spacing. This length is the ``width'' of the fluctuations of the particles
as shown in Figure~\ref{widthfluct}.
Note the difference between the two lengths $l_{\perp}$ and
$L_{\rm loc}$. They are related through
\cite{giamarchi_columnar_variat}
\begin{equation} \label{locrad}
l_{\perp}^2 = \frac{\alpha^2}{\pi^2} (\hbar \kbar)
\ln (L_{\rm loc}/a)
\end{equation}
At the transition one expects that $L_{\rm loc}$ diverges as
$L_{\rm loc} \sim \exp(b/(K_c-K)^\alpha)$ with $\alpha = 1/2$ from
the RG  (see equation (\ref{eq:lloc-bkt})). Thus relation
(\ref{locrad}) predicts that:
\begin{equation}
l_{\perp} \sim \frac{1}{(K_c-K)^{\alpha/2}}
\end{equation}
diverges as a power law, which could be measured in numerical
simulations.

\begin{figure}
\centerline{\psfig{file=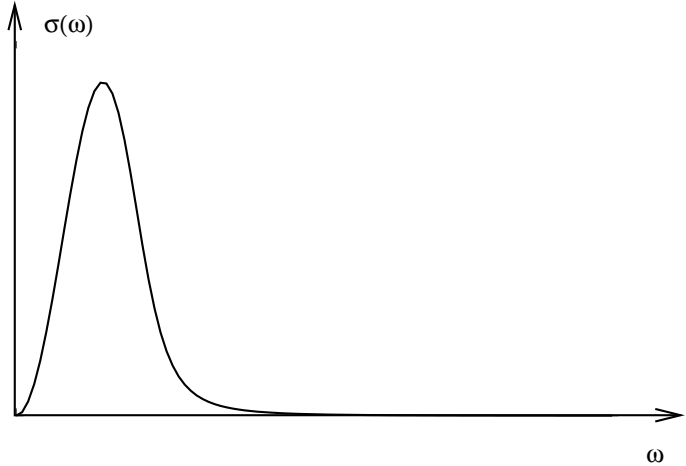,angle=0,width=\figwidth}}
\caption{The conductivity of the one dimensional Anderson
insulator in the limit $K \to 0, \hbar \to 0, \kbar=\frac K
\hbar$ fixed.} \label{fig:condanderson}
\end{figure}

Let us now look at the conductivity, for which we need $I(\omega_n)$.
If we call $I'(\omega)$ and $I''(\omega)$ the real and imaginary parts
of the analytic continuation of $I(\omega_n)$, then the conductivity
is given by
\begin{equation} \label{eq:condcomp}
\sigma(\omega) = \frac{\omega I'' + \omega (-\omega^2 + I' + \Sigma_1)}
{(-\omega^2 + \Sigma_1 + I')^2 +(I'')^2}
\end{equation}
It is easy to see that $\Re\sigma(\omega)$ goes to zero at zero
frequency and thus the phase is indeed localized. But {\it
because} of the analytic continuation, the existence or not of a
gap in the optical conductivity is {\it not} linked to the
existence of a ``mass'' $\Sigma_1$ but to whether $I''$ is
nonzero at small frequencies or not. The equation for
$I(\omega_n)$ takes a particularly simple form in the limit
$K,\hbar \to 0$ while keeping $\overline{K}$ fixed. In this
limit, we can write
\begin{equation}
\label{eq:scalingivaria}
I(\omega_n)=\Sigma_1 f(\frac{\omega_n}{\sqrt{\pi \kbar \Sigma_1}}),
\end{equation}
where the scaling function $f$ satisfies:
\begin{equation}
\label{eq:fukuleevaria}
f(x)=2 \left[ 1- \frac 1 {\sqrt{1+x^2+f(x)}}\right].
\end{equation}
It can be shown that for $\omega \to 0$
\begin{equation}
\Re\sigma(\omega) \sim
\omega^2
\end{equation}
In a similar way (\ref{eq:condcomp}) gives
\begin{equation}
\Im\sigma(\omega) \sim \omega/\Sigma_1
\end{equation}
Such behavior is in agreement with exact results in one dimension
\cite{berezinskii_conductivity_log,abrikosov_rhyzkin} up to
logarithmic prefactors. At high frequency, one can show that
$\sigma(\omega)\sim \omega^{2K-4}$ in agreement with the RG
result. The resulting conductivity is plotted on
Figure~\ref{fig:condanderson}.

\subsection{The fine prints} \label{eq:fineprints}

The variational method is thus an extremely efficient method for
this type of disordered problems. It allows to get most of the
physical properties in the localized phase, and give the
qualitative features of the transition and the delocalized phase.
Its physics is very similar to the one described for point like
disorder in section~\ref{classical}. The variational method
determines the ``static'' solution for the mode $\omega_n=0$ (but
by taking into account the effects of all modes) without being
restricted to a single Gaussian. It can then correctly compute the
fluctuations (both in $q$ and $\omega_n$) around this solution,
contrarily to the approximate method of section~\ref{scha} which
was impeded by the lack of knowledge of the static solution.

In $d=1$ an additional remarkable property can be seen. Although
most of the properties are independent of the value of $u_c$, the
conductivity is strongly dependent on it. As for the static
solution one has continuous replica symmetry breaking for $d>2$
which goes continuously to the one step solution in $d=2$. In
these dimensions taking the value of $u_c$ out of the variational
equations gives the correct conductivity (no gap in
$\sigma(\omega)$). In $d=1$, the $u_c$ obtained by minimizing the
free energy would give an {\it incorrect} (gapped) conductivity.
Another way to determine $u_c$ is to use the marginality
condition \cite{giamarchi_columnar_variat} that corresponds to the
instability of the replicon mode, similar to the condition
(\ref{replicon}). This condition, more dynamical in nature,
coincides with the free energy value of $u_c$ for $d \geq 2$ but
is different in $d=1$. One can check that the marginality
condition always gives the correct conductivity. Some arguments
for why it is so were given in \cite{giamarchi_columnar_variat},
but this point is not yet fully understood. This phenomenon has
since been found to occur in other systems
\cite{georges_quantum_spinglass}.

\begin{figure}
\centerline{\psfig{file=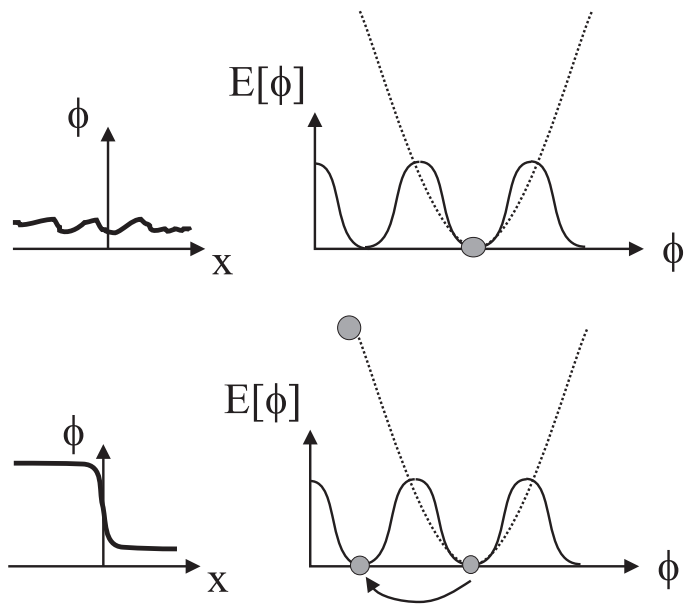,angle=0,width=\figwidth}}
\caption{\label{fig:soliton}The gaussian approximation treats
correctly the excitations where the phase stays in one of the
minima of the cosine. They contribute to the a. c. conductivity at
$T=0$. Soliton excitations bring the phase from one minimum to
another. With the $\cos(\phi)$ term these excitations only cost
energy at the kink. When the potential is replaced by the
quadratic approximation, the energy cost of the ``wrong'' minimim
is very high, hence a poor result for the physical results. These
excitations dominate the d.c. conductivity for $T>0$.}
\end{figure}

The success of the variational method is obviously linked to the fact
that we are looking at small ``quadratic'' fluctuations around a
certain (in our case highly disordered) solution. In this case the
replacement
\begin{equation} \label{eq:approx}
-\cos(\phi) \to \frac12 \phi^2
\end{equation}
is very reasonable. Such an approximation is very good to compute
correlation functions of the variable $\phi$. This is the case for
the density-density correlation and the optical conductivity.
What is missing in the approximation (\ref{eq:approx}) are the
solitons that go from one minimum of the potential to another
minimum. The energy cost of such excitations is grossly
overestimated by the variational method as is shown on
Figure~\ref{fig:soliton}.

This has important consequences for the calculation of
correlation functions that involve the soliton creation operator
$e^{i \theta}$. These correlation functions are found incorrectly
to decay exponentially with distance at equal time  and to be
zero at unequal time. As a result, the variational method does
not allow the calculation of Fermion Green's function (they
involve the operators $e^{i (\theta\pm\phi)}$) nor
superconducting correlations (which involve the operator
$e^{2i\theta}$). In the case of a disordered XXX spin chain, the
situation is even worse, since $S_x\propto \cos(\theta)$, whereas
$S_z \sim -\frac{\partial_x \phi}{\pi} + (-)^{x/a} \cos 2\phi$.
Therefore, even in the presence of a disorder that preserves
SU(2) symmetry (such as a random bond disorder), the variational
method would lead to a spurious breaking of rotational symmetry.
It would also give poor results for systems that include both the
$\theta$ and the $\phi$ field in the Hamiltonian. Examples of
such theories include disordered Hubbard ladders
\cite{orignac_2chain_long} , disordered spin
ladders\cite{orignac_2spinchains} or XXX spin chain in a random
fields \cite{doty_xxz}.

\begin{figure}
\centerline{\psfig{file=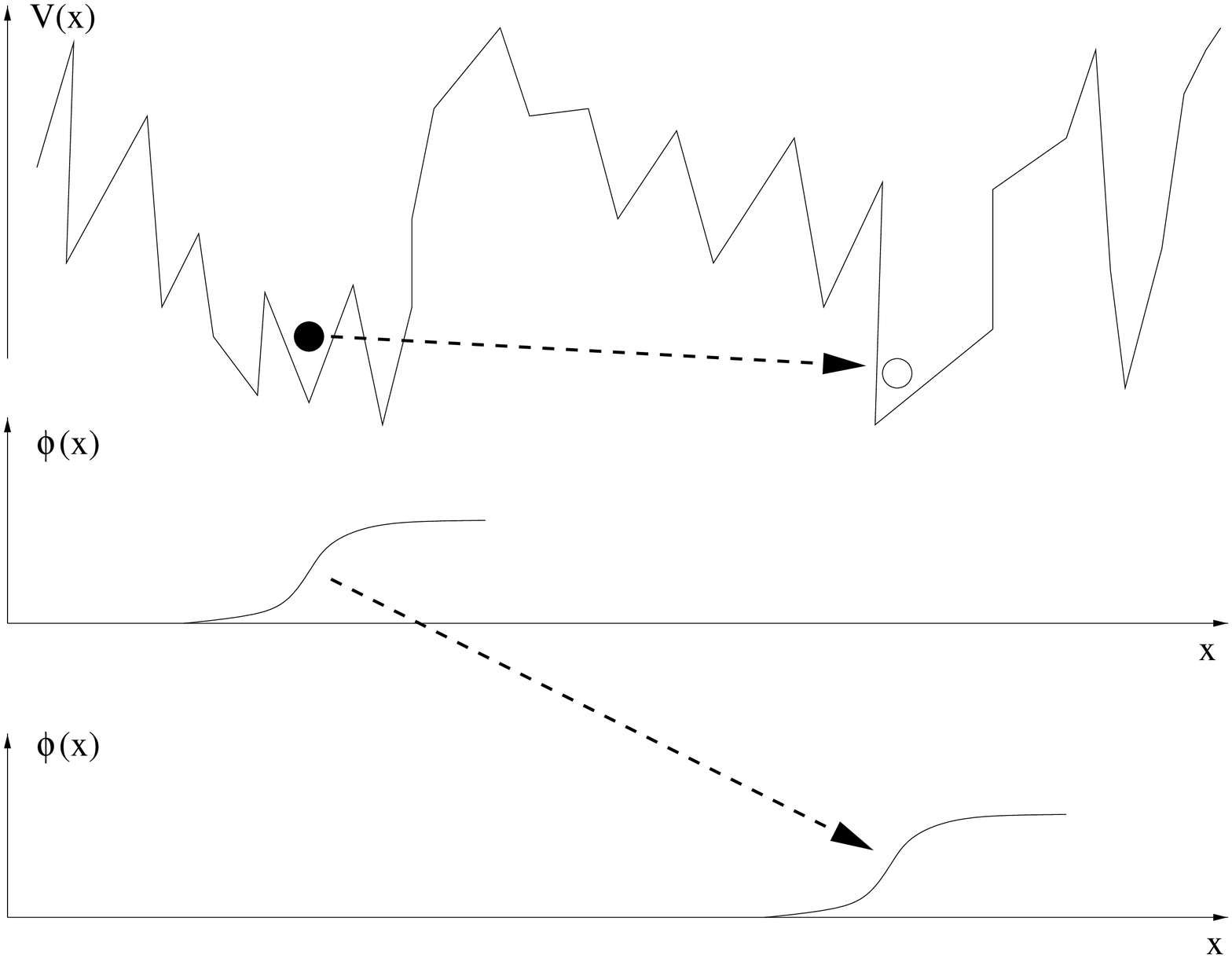,angle=0,width=\figwidth}}
\caption{\label{fig:mott} Hopping of charges that leads to a
finite d.c. conductivity. This corresponds to soliton-type
excitations in the phase variable $\phi$.}
\end{figure}

The mishandling of soliton excitations also limits our knowledge
of the transport properties at finite temperatures. Indeed the
optical conductivity does not correspond to transport of charge
but charge oscillations around the equilibrium positions, it is
thus well described by our harmonic approximation as shown in
Figure~\ref{fig:soliton}. On the contrary transport at finite $T$
involves real charge displacements. Since $\rho(x) \sim
\nabla\phi$, displacing a charge amounts to make a solitonic
excitation in the field $\phi$ as shown in Figure~\ref{fig:mott}.
Indeed Mott's arguments, to compute the conductivity of
noninteracting electrons in presence of phonons $\sigma(T)\sim
\exp\left[-\left(\frac{T_0} T\right)^{\frac 1 {d+1}}\right]$ is
strongly reminiscent of an instanton calculation as shown in
Figure~\ref{fig:mott}.
A similar argument applies to Efros and Schklovskii
calculations\cite{efros_coulomb_gap}. An extension of the
variational approach to finite temperatures would indeed still
lead to $\sigma(T,\omega=0)=0$ proof that it is missing the
excitations that are important at finite temperature.
Unfortunately no way to treat such solitons has been found at
present despite some attempts \cite{lee_larkin_conductivite_cdw}.

Despite these limitations, the variational method is up to now the only
analytical method giving information for such localized systems in the
localized phase. As with all variational approaches, some physical
insight in the properties of the system under consideration is needed
to determine whether the method as any chance of success. Clearly, one
must apply this method only to systems that can be reasonably well
understood qualitatively from their classical action. Fortunately many
systems fall in this category, and we examine some of those in the
following.

\subsection{Higher dimension: electronic crystals and classical systems}

\begin{figure}
\centerline{\psfig{file=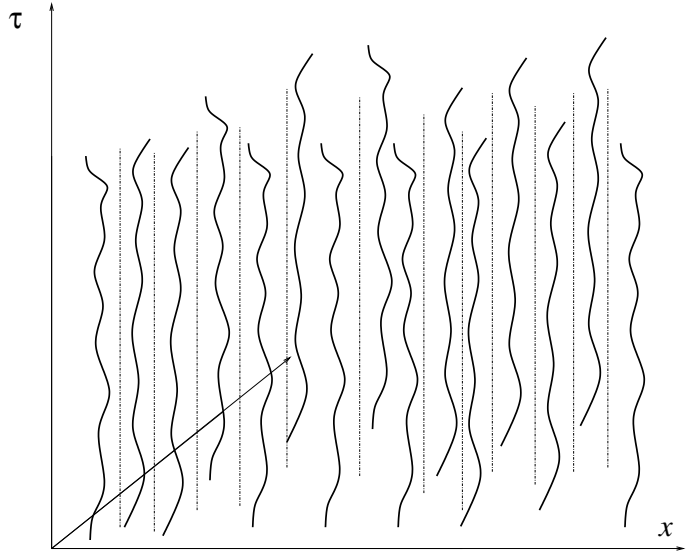,angle=0,width=\figwidth}}
\caption{The action (\ref{eq:repaction}) in $d=2$ describes
elastic lines in the presence of columnar pins.} \label{fig:lines}
\end{figure}

First the GVM can be used to study classical systems using the
standard mapping $\tau \to z$. The action (\ref{eq:repaction})
and its extension to higher spatial dimension describes elastic
objects (lines in this case as shown in Figure~\ref{fig:lines})
pinned by columnar (i.e. time or $z$ independent) defects. This
situation is realized for example in vortex in type II
superconductors irradiated by heavy ions (creating the linear
track of disorder). This system in $2+1$ dimension is equivalent
to a $d=2$ quantum bose system in presence of pins. In a similar
way than in $d=1$ (see section~\ref{sec:quantumvaria}) such
system has a pinned phase (the Bose glass)
\cite{fisher_bosons_scaling,nelson_columnar_long}. The
variational method can be used to describe the Bose glass phase
\cite{giamarchi_columnar_variat}. However contrarily to $d=1$ it
cannot be used to go to the superfluid regime since to describe a
two dimensional ``melting'' of the Bose glass phase dislocations
are important (no dislocations exist in $d=1$) and for reasons
explained above the GVM overestimates the energy cost of
topological excitations. Another way to say it, is that in $d
>1$ we loose the elastic description
(\ref{eq:density-bosonized-general}) of the Fermion or Boson
operators. The GVM can thus only be used in phases where the
particles are localized so that some elastic description can
again be used.
We can thus use the variational method in higher dimensions to
study electronic crystals. This included Charge density wave, but
also the two dimensional Wigner crystal of electrons. In such a
phase the electrons are confined by their repulsion (and a in
some systems a magnetic field). An elastic description can be
used. Some level of quantumness is hidden in the elastic
parameters (``size'' of the particles, quantization of the phonon
modes of the crystal). For such systems the calculation of the
optical conductivity is particularly useful since it is one of
the few probes of such systems. Since the physics of such systems
would deserve a review of its own we will not dwell further on it
here but refer the reader to
\cite{chitra_wigner_hall,chitra_wigner_long} for details.

\section{Commensurate systems} \label{sec:comm}

When the filling of the fermion system is commensurate the physics
discussed above is modified in various ways since the backward
scattering on disorder becomes real. If the forward scattering still
exists, not much is changed. Two special cases will thus occur: (i) the
forward scattering is absent. This occurs because of a symmetry of the system. This is the
case for example for fermions at half filling with a random hopping or
for spin chains with random exchange. (ii) The commensurate potential
(either due to the lattice or due to electron-electron interaction)
would open a gap. There will be a competition between Mott physics
wanting to get a commensurate (gapped) insulator and the disorder that
would like to destroy such a gap (push the system locally away from
commensurability).

\subsection{The peculiar random exchange}

\begin{figure}
\centerline{\psfig{file=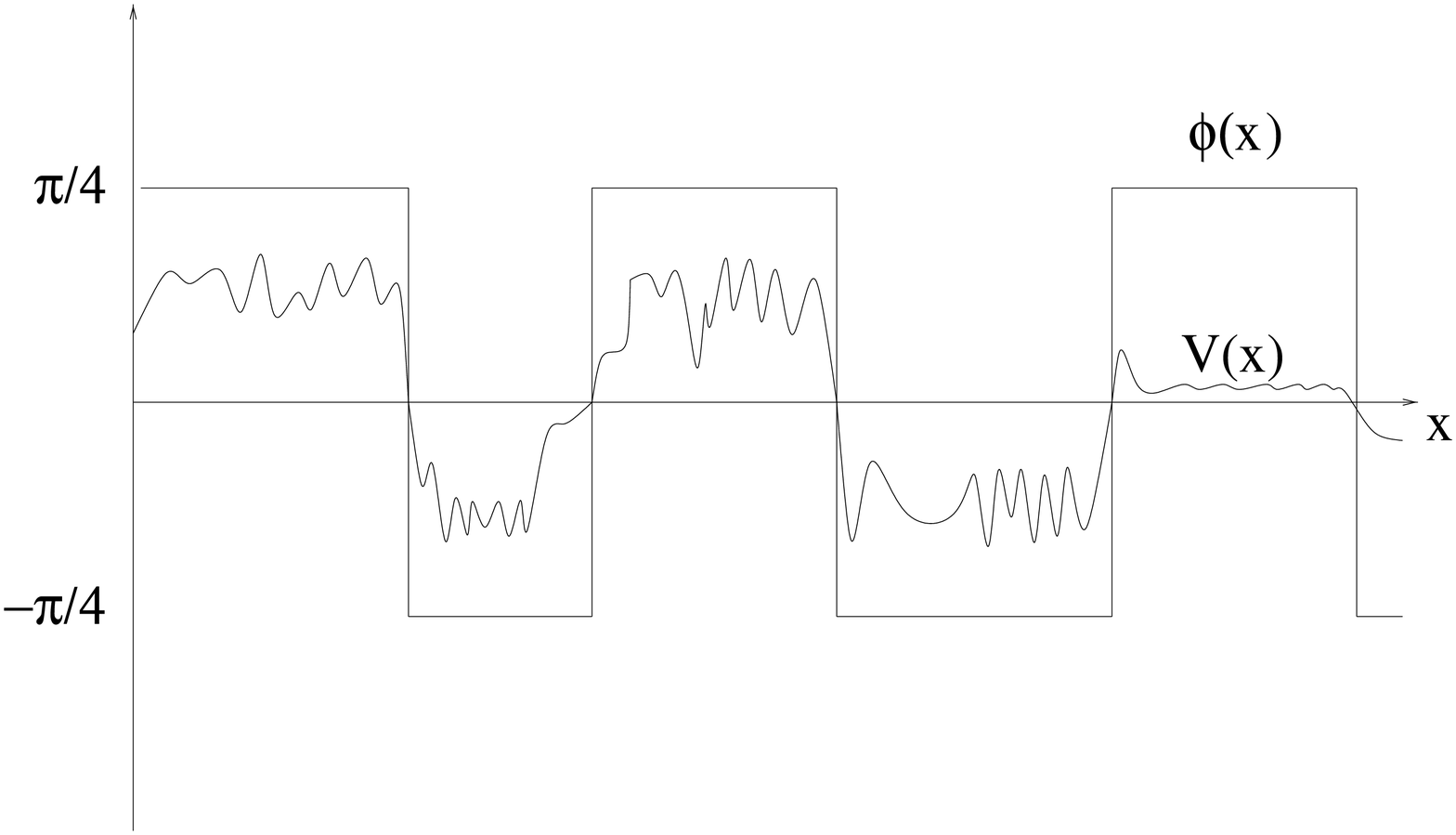,angle=0,width=\figwidth}}
\caption{The profile of $\phi(x)$ for a given realization of disorder
$V(x)$. As one can see, $\phi(x)$ is only sensitive to the \emph{sign}
of $V(x)$ and not to its amplitude. As a result, $\phi$ has kinks each
time $V(x)$ goes to zero.}
\label{fig:kinks}
\end{figure}

For electrons at half filling with a random exchange the forward
scattering does not exist and the disorder term is simply
\begin{eqnarray}\label{eq:random-exchange}
H &=& \int dx V(x)i[\psi^\dagger_+\psi_-  - {\rm H. c.}] \\
&=& - \int dx \frac{V(x)}{(2\pi\alpha)}\sin(2\phi(x))
\end{eqnarray}
Although this seems very similar to (\ref{eq:incaction}) one easily sees the
difference on Figure~\ref{fig:kinks}. Contrarily to normal disorder
where $\phi$
follows the random {\it phase} of the random potential, here $\phi=\pm
\pi/4$ depending on the sign of the potential. Thus $\phi$ is nearly
gapped but for its kinks. The low energy properties will thus be
dominated by the kinks in $\phi$. Such kink structure between the
doubly degenerate minima makes it unlikely that the GVM can be used for
this problem.

In order to get an idea of the physics let use examine the case
of free fermions $K=1$. For $E=0$ one can easily construct the
eigenstates by solving
\begin{eqnarray}
\partial \psi_+ &-&  V(x)\psi_-(x) = 0\\
-\partial \psi_- &+& i V(x)\psi_+(x) = 0
\end{eqnarray}
to obtain the (unnormalized)
\begin{equation}
\psi_+(x) \sim \psi_-(x) \sim e^{-\int_0^x dy V(y)}
\end{equation}
which obviously decays as
\begin{equation}
\psi_\pm(x) \sim e^{-\sqrt{x}}
\end{equation}
If one wants to view such a state as an exponentially localized
state this means that the localization length diverges when $E\to
0$. The divergence of the localization length is of course the
signature that the physical properties will be quite different
than the ones for the ``normal'' disorder studied in the previous
sections. For the non-interacting case they have been studied
using a variety of techniques such as replicas, Berezinskii
methods, supersymmetry etc.
\cite{gogolin_disorder_revue,comtet_replica_dirac,gogolin_random_mass,%
balents_random_dirac}. The density of states diverge at the Fermi
level as\cite{eggarter_random_hopping}
\begin{equation}\label{eq:singular_dos}
\rho(\epsilon) \sim \frac 1{\epsilon \log(1/\epsilon)^{3}}
\end{equation}
Quite astonishingly if one considers that all states are still
localized the d.c. conductivity is now {\it finite}
\cite{gogolin_disorder_revue,damle_random_exchange}. This is
surprising since it seems to violate the Mott phenomenological
derivation. The conductivity is proportional to the absorbed
power. To make a transition between two localized states one need
one occupied state, one empty one separated by an energy $\hbar
\omega$. The number of such states is $\rho(\epsilon=0) \hbar
\omega$. The absorbed power (and the conductivity) is thus (up to
log correction)
\begin{equation}
\sigma(\omega) \propto \rho(\epsilon=0) (\hbar \omega)^2
\end{equation}
correctly giving back for a constant density of states at the Fermi
level the $\omega^2$ dependence of section~\ref{sec:quantumvaria}.
One would thus
naively have expected a $\sigma(\omega) \sim \omega$ in the case of a
singular density of state (\ref{eq:singular_dos}).
A qualitative understanding of the behavior of the d.c. conductivity
is thus still lacking.

When interactions are included, the problem becomes more
difficult to tackle. Going back to the spin chain version of
(\ref{eq:random-exchange}) a real space renormalization procedure
has been introduced. This procedure works beautifully and allows
the calculation of most correlation functions. We refer the
reader to \cite{fisher_random_transverse,fisher_rtfim} for
details. As for the normal disorder (see section~\ref{sec:rg})
interaction were found to be irrelevant at the disordered fixed
point. The divergence of the localization length also changes
drastically the correlation functions. In particular the spin spin
correlation functions now decay as power law instead of being
exponential. It would be extremely interesting to have an
equivalent derivation of this real space RG directly in the boson
representation.

\subsection{Mott versus Anderson}

A particularly
interesting situation occurs when the non-disordered system
possesses a gap. In that case the competition between this gap and
the disorder is non trivial. This arises in a large number of systems
such as disordered Mott insulators
\cite{shankar_spinless_conductivite,fujimoto_mott+disorder_1ch,mori_scba,%
paalanen_silicon_phosphorus},
systems with external (Peierls or spin-Peierls
systems \cite{grenier_cugesio3}) or internal commensurate potential
(ladders or spin ladders
\cite{orignac_ladder_disorder,azuma_zinc_doping,carter_lacasrcuo,%
fujimoto_mott+disorder_2ch}, disordered spin 1 chains
\cite{monthus_spin1_ranexchange,kawakami_dopeds=1}).
It is physically simple to see that in order to destroy the gap one
needs a disorder comparable to the gap (although in some specific cases
Imry-Ma effects can destroy the gap for infinitesimal disorder
\cite{shankar_spinless_conductivite}).  This makes
the complete description of the gap closure and of the
physics of the resulting phases is extremely difficult with the usual
analytic techniques such as the perturbative renormalization group described
in section~\ref{sec:rg}, due to the absence of a weak coupling fixed point.
Indeed both the commensurate Mott insulator
(which can be described by a sine-Gordon
Hamiltonian \cite{giamarchi_mott_shortrev}) and the disordered Anderson
insulator correspond to strong coupling fixed points.

To be concrete, but keep the technical details to a minimum let us
consider again spinless fermions. The commensurability can be described
by adding to the Hamiltonian (\ref{eq:hambos}) a periodic potential at
the Fermi level
\begin{equation}
H_{\rm com} = g\int dx (\psi^\dagger_+\psi_-+\psi^\dagger_-\psi_+)
\end{equation}
which becomes in the boson representation
\begin{equation} \label{eq:combos}
H_{\rm com} = \int dx \frac{g}{\pi\alpha\hbar} \cos 2\phi
\end{equation}
The reason to consider here a periodic potential is specific to
pathologies associated to the Mott gap for spinless fermions\cite{shankar_spinless_conductivite}. Similar
results are expected for a Mott insulator (which is due to the $4k_F$
potential of the lattice).

If this system is studied by RG then one is faced with the competition
of two strong coupling fixed points since
\begin{eqnarray}
\frac{d g}{d l} &=& (2-K)g \\
\frac{d \D}{d l} &=& (3-2K) \D
\end{eqnarray}
Using the usual qualitative argument consisting in taking the most
divergent operator, it was concluded
\cite{fujimoto_mott+disorder_1ch} that if disorder reaches strong
coupling ($\D(l^*)=1>g(l^*)$) first, we will be in the Anderson
phase, with a localization length $l_0=ae^{l^*} \sim \left(\frac
1 \D \right)^{\frac 1 {3-2K}}$, whereas if the commensurate
potential reaches strong coupling first, we will be in the Mott
phase with a correlation length (or soliton size) $d \sim
\left(\frac 1 g \right)^{\frac 1 {2-K}}$. The phase transition
between the two phases occurs for $l_0 \sim d$. This picture
relies on the important assumption that there is no other stable
fixed point than the Mott insulator and the Anderson insulator.
Even if it was so, it would not be possible to determine the
phase boundaries, nor determine what type of critical point
separate the Mott and the Anderson insulator. In order to make
progress on these issues and obtain a more complete picture of
dirty one-dimensional Mott insulators, one needs to solve the
problem non-perturbatively \cite{orignac_mg_short}. The methods of
section (\ref{sec:quantumvaria}) are well adapted since the
problem can be cast in the sine-Gordon form.

\subsection{Variational approach}\label{sec:commvaria}

For the Mott versus Anderson problem, the variational action
reads:
\begin{eqnarray}\label{eq:pp+cdbreplica}
\frac{S_{\rm rep. }}{\hbar}&=&\sum_a \left[ \int \frac{dx d\tau}{2\pi K}
\left(v (\partial_x \phi_a)^2 + \frac{(\partial_\tau \phi_a)^2}{v}\right)
-\frac {g}{\pi\alpha \hbar} \int dx d\tau \cos 2 \phi_a \right]
\nonumber \\
&-& \frac {\D}{(2\pi\alpha \hbar)^2} \sum_{a,b} \int dx
\int_0^\beta d\tau d\tau' \cos
\left(2(\phi_a(x,\tau)-\phi_b(x,\tau'))\right)
\end{eqnarray}
The term $\cos 2\phi_a$ in (\ref{eq:pp+cdbreplica}) is responsible
for the opening of a gap. We search for a saddle point with a
form of the variational connected Green's Function slightly
generalized with respect to (\ref{eq:gcrsb}):
\begin{equation} \label{eq:solution}
v G_c^{-1}(q,\omega_n) = \frac1{\pi  \overline{K}} ( \omega_n^2 + v^2
q^2) + m^2 + \Sigma_1(1-\delta_{n,0}) + I(\omega_n)
\end{equation}
Where the parameter $\Sigma_1$ and the function  $I(\omega_n)$
satisfy the equations (\ref{eq:Sigmarsb}) and
(\ref{eq:iomegarsb}). We give here the main steps of the solution
and refer the reader to
\cite{orignac_mg_short,giamarchi_commensurate_long} for more
details. The variational self-energy satisfies the equation
(\ref{eq:sigmaursb}). Finally, $m$ satisfies the equation:
\begin{equation}\label{eq:mrsb}
m^2= \frac{4 g  v }{\pi \alpha
}e^{-2\hbar\tilde{G}(0,0)}
\end{equation}
The important physical quantities are simply given such as
conductivity and compressibility are simply given by
(\ref{conduc}) and (\ref{replica_compressiblility}).

\begin{figure}
\centerline{\psfig{file=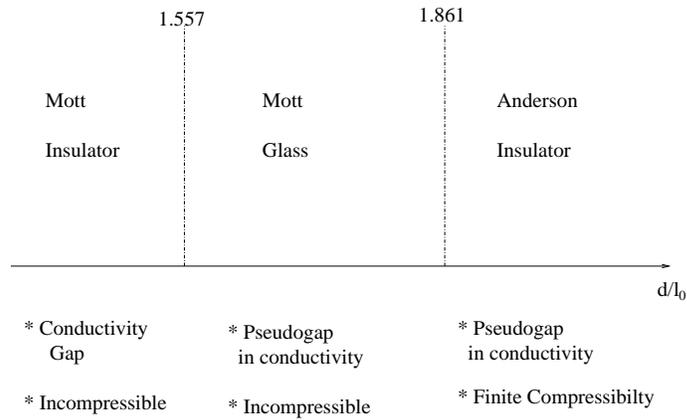,angle=0,width=\figwidth}}
\caption{The phase diagram of a system with commensurability and
disorder. $d$ is the soliton size, $l_0$ the localization (or pinning
length)}
\label{fig:phases}
\end{figure}

\begin{figure}
 \centerline{\psfig{file=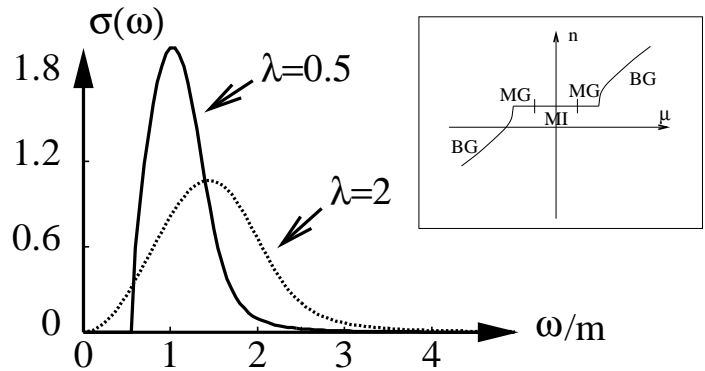,angle=0,width=\figwidth}}
  \caption{\label{fig:conduc} Conductivity in the Mott Insulator (solid line)
   and Mott Glass phases (dashed line). Insert: density $n$ versus the chemical
  potential $\mu$. $\lambda$ is defined in the text.}
\end{figure}

Since we expect the physics to be continuous for small enough $K$
(i.e. repulsive enough interactions), one can gain considerable
insight by considering (see sec. \ref{scha}) the classical limit
$\hbar \to 0$, $K\to 0$ keeping $\overline{K}=K/\hbar$ fixed. In
this limit one can solve analytically the saddle point equations
(\ref{eq:gcrsb})--(\ref{eq:sigmaursb}),
(\ref{eq:solution})--(\ref{eq:mrsb})  and compute $m$, $\Sigma_1$
and $I(\omega_n)$. The resulting phase diagram is parameterized
with two physical lengths (for $K\to 0$): The  correlation length
(or soliton size) of the pure gapped phase
\begin{equation}
d=\left(\frac{4 g \overline{K}}{ \alpha v}\right)^{-1/2}
\end{equation}
and the localization (or pinning)  in the absence of
commensurability length
\begin{equation}
l_0=\left(\frac {(\alpha v)^2}{16 \D
\overline{K}^2}\right)^{1/3}
\end{equation}
Contrarily to the naive direct transition predicted by the
extrapolation of the RG, we find {\it three} phases
\cite{orignac_mg_short} as shown in Figure~\ref{fig:phases}.
Their main characteristics in term of conductivity and
compressibility are summarized in Figure~\ref{fig:conduc}.

{\it Mott insulator}

At weak disorder we find a replica symmetric solution with
$\Sigma_1 = 0$ but with $m \ne 0$ ($m$ depends on the disorder).
 $m$ is given by the simple equation:
\begin{equation} \label{sigmaomega=0RS}
m^2=\frac {4 g  v}{\pi \alpha} \exp
\left[ -\frac{ \D  \overline{K}^{1/2} }{ \alpha^2 \pi^{3/2} v^{1/2}
m^3} \right].
\end{equation}
It is convenient to work with the physical lengths $d$ and $l_0$
and the correlation length $\xi$ where:
\begin{equation}
\xi^2= \frac{v^2}{(\pi \overline{K} m^2)}.
\end{equation}
One can then rewrite (\ref{sigmaomega=0RS}) as:
\begin{equation}\label{eq:mreduced}
\frac 1 {\xi^2}=\frac 1 { d^2}\exp\left[-\frac 1 {16} \left(
\frac \xi  {l_0}\right)^3\right].
\end{equation}
For $l_0/d >\frac 1 2 \left( \frac {3e} 4 \right)^{1/3}$,
(\ref{eq:mreduced}) has a single physical solution. For $l_0/d
<\frac 1 2 \left( \frac {3e} 4 \right)^{1/3}$,
(\ref{eq:mreduced}) has no solution, which means that the Mott
insulator becomes unstable. In fact due to another contraint (see
below) the Mott Insulator becomes unstable at an even smaller
disorder.

Having  $m \ne 0$ leads to zero compressibility $\kappa = 0$.
Disorder reduces the gap created by the commensurate potential
and thus  increases $\xi$ compared to the pure case. Let us now
examine the equation for $I(\omega_n)$ giving the transport
properties. An expansion around $\hbar=0$ in equation
(\ref{eq:iomegarsb}) gives the self-consistent equation for
$I(\omega_n)$:
\begin{equation}\label{eq:equationrsi}
I(\omega_n)=\frac{ 8 \D   v}{ (\pi \alpha)^2}\int_0^{\beta \hbar}
G_c(x=0,\tau)(1-\cos(\omega_n \tau)) d\tau
\end{equation}
Introducing the scaling form (to be contrasted with (\ref{eq:scalingivaria})):
\begin{equation}\label{eq:iomegareduced}
I(\omega_n)= m^2 f\left(\frac{\omega_n}{\sqrt{\pi \overline{K}}m
}\right),
\end{equation}
(\ref{eq:equationrsi}) can be recast in the form:
\begin{equation} \label{equation-for-f}
f(x) = \lambda \left[ 1-\frac 1 {\sqrt{1+x^2+f(x)}}\right]
\end{equation}
where:
\begin{equation}\label{eq:def-lambda}
\lambda=\frac{4 \D \overline{K}^{1/2} v }{ \pi^{3/2} \alpha^2
  m^3}=\frac 1 {4} \left(\frac \xi {l_0}\right)^{3}
\end{equation}
Let us note that for $\lambda=2$, the equation
(\ref{equation-for-f}) reduces to the equation
(\ref{eq:fukuleevaria}). For $\lambda>2$ (\ref{equation-for-f})
has no physical solution. Using (\ref{eq:def-lambda}),  this
condition becomes:
\begin{equation}\label{eq:conditionmarginal}
 \frac {l_0}{d}< \frac 1 2 e^{\frac 1 {4}}
\end{equation}
For $\lambda <2$, there is a physical solution of
(\ref{equation-for-f}) such that  $\lim_{x \to \pm \infty} f(x)=
1+\lambda$  and for $x \ll 1$ , $f(x)=1+\alpha x^2 +o(x^2)$ with
$\alpha=\lambda/(2-\lambda)$. The conductivity of the Mott
insulator can be obtained from $f$ in the form:
\begin{equation}\label{conductivite-f}
\sigma(\omega)=\frac{ \xi \overline{K} }{\pi}\frac{\imath
x}{(1+f(\imath x)-x^2)}
\end{equation}
where $x=\omega/\omega^*$ and $\omega^*=v/\xi$ is the characteristic
frequency associated with the correlation length $\xi$.
The conductivity $\sigma(\omega)$ is zero if:
\begin{equation} \label{criticalx}
\omega<\omega_c=\omega^*\sqrt{1+\lambda-3\left(\frac \lambda 2 \right)^{2/3}}
\end{equation}

\begin{figure}
\centerline{\psfig{file=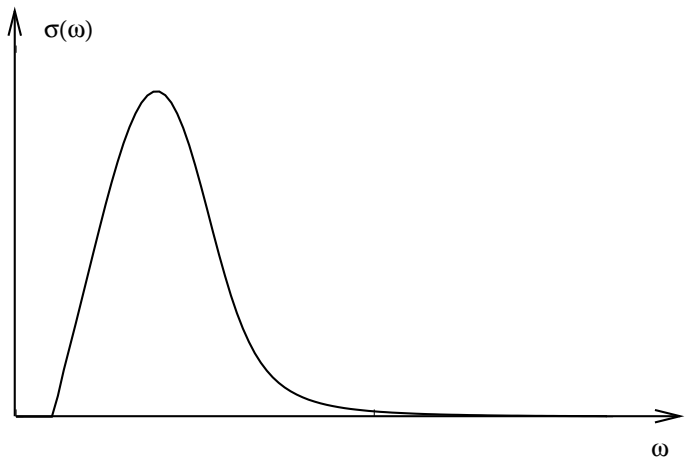,angle=0,width=\figwidth}}
\caption{The frequency dependant conductivity of the Mott Insulator}
\label{fig:condmi}
\end{figure}

\begin{figure}
\centerline{\psfig{file=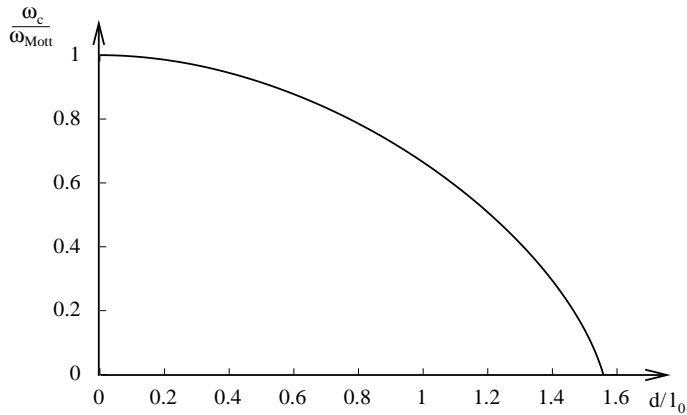,angle=0,width=\figwidth}}
\caption{The dependance of the conductivity gap with disorder
strength in the Mott insulator} \label{fig:condgap}
\end{figure}

The replica symmetric phase with $m\ne 0$ has thus a conductivity
gap $\hbar\omega_c$ and can be assimilated to a Mott insulator
(MI). The behavior of the conductivity is plotted on figure
(\ref{fig:condmi}). Close to the threshold, one has
$\sigma(\omega)\sim (\omega-\omega_c)^{1/2}$.
However the gap in the conductivity $\hbar \omega_c$ decreases
when disorder increases, and closes for $d/l_0=2e^{-1/4}\simeq
1.557 $ (see Figure \ref{fig:condgap}). For $d/l_0 \to
2e^{-1/4}$, the conductivity gap vanishes linearly with $\frac d
{l_0}$.
For $d/l_0>2e^{-1/4}$  the MI solution becomes unphysical
\emph{even though} the mass $m$ remains finite at this transition
point, i.e. the system remains incompressible. For stronger
disorder one must break replica symmetry, as for the pure
disorder case of section~\ref{sec:quantumvaria}. Here, however
\emph{two} possibilities arise depending on whether the saddle
point allows for $m \ne 0$ or not. In the presence of a breaking
of replica symmetry, one extra equation is needed to determine
the breakpoint. As in the case of the Anderson insulator, such
equation is provided by the marginality of the replicon condition
discussed in section~\ref{eq:fineprints}. Two phases exists:

{\it The Anderson Glass}

For large disorder compared to the commensurate potential $d/l_0
> 1.861$, $m=0$ is the only saddle point solution. The saddle
point equations then reduce to those
(\ref{eq:gcrsb})--(\ref{eq:sigmaursb}) . Thus, we recover the
Anderson glass with interactions of section
(\ref{sec:quantumvaria}). As we have seen in section
(\ref{sec:quantumvaria}), in such phase the conductivity starts
as $\sigma(\omega) \sim \omega^2$ showing no gap and the
compressibility is finite.

In the Anderson glass phase, the disorder washes out completely
the commensurate potential.  The MI and the AG were the only two
phases accessible by renormalization techniques
\cite{fujimoto_mott+disorder_1ch}. Within the replica variational
formalism however, we find that an intermediate phase exists
between them.

{\it The Mott Glass}

For intermediate disorder $2e^{-1/4} < d/l_0 < 1.861$ a phase
with \emph{both} $\Sigma_1 \ne 0$ and  $m \ne 0$ is obtained. We
shall call this phase the {\it Mott Glass} (MG). We shall not
discuss in full detail the one-step solution of the saddle point
equation here. we will rather stress the salient features of our
solution. First, as a result of the marginality of replicon mode
condition, $m^2 + \Sigma_1$ remains constant in the MI and MG as
disorder strength is increased
\cite{orignac_mg_short,giamarchi_commensurate_long}. In the MG
phase, $I(\omega_n)$ is still of the form
(\ref{eq:iomegareduced}) but $m$ is replaced by
$\sqrt{m^2+\Sigma_1}$. The reduced self-energy $f(x)$ satifies
(\ref{equation-for-f}) but with $\lambda=2$ in the whole Mott
Glass phase. This implies that (see (\ref{eq:fukuleevaria})) that
the a.c. conductivity of the Mott Glass is identical to the one
of an Anderson glass. However, since $m \ne 0$ in the MG, the
system is {\it incompressible} ($\kappa = 0$) like a Mott
Insulator. Thus, the Mott Glass is a new glassy phase (since it
has Replica Symmetry Breaking) with characteristics intermediate
between those of an Anderson Insulator and those of a Mott
Insulator.

\subsection{Physical discussion}

The existence of a phase with a compressibility gap but no
conductivity gap is quite remarkable since by analogy with
noninteracting electrons\cite{mori_scba} one is tempted to
associate a zero compressibility to the absence of available
states at the Fermi level and hence to a gap in the conductivity
as well. Our solution shows this is not the case, when
interactions are turned on  the excitations that consists in
adding one particle (the important ones for the compressibility)
become  quite different from the particle hole excitations that
dominate the conductivity. A similar situation is obtained in the
case of the one dimensional Wigner crystal
\cite{schulz_wigner_1d}, which has the conductivity of a perfect
1d metal, $\sigma(\omega) \propto \delta(\omega)$ but a zero
compressiblity since $\chi=\lim_{q \to 0} \frac 1 {\ln q}$. This
argument suggest that the difference in one-particle and
two-particle properties is a consequence of the strong repulsion
in the system.

\begin{figure}
\centerline{\psfig{file=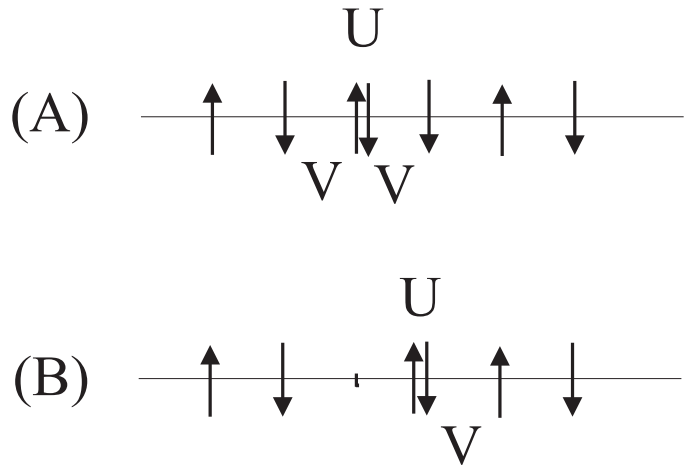,angle=0,width=\figwidth}}
\caption{\label{fig:exciton} (A) Energy needed to add one
particle. (B) Energy needed to make the particle hole excitations
entering in the optical conductivity. Because of excitonic
processes, when disorder is added the gap in the optical
conductivity can close before the single particle gap. This
phenonenon, leading to the MG phase, should occur regardless of
the dimension.}
\end{figure}

In addition to the variational method itself the Mott glass phase
can also be obtained by two other independent methods. Higher
dimensional extensions of the present problem, similar to the one
made in section~\ref{classical} can be treated around four spatial
dimensions using a $d=4-\epsilon$ functional renormalization
group method (totally different form the $d=2$ RG). Such study
confirms \cite{orignac_mg_short,giamarchi_commensurate_long} the
existence of the intermediate Mott glass phase. One can also
analyse (\ref{eq:pp+cdbreplica}) for zero kinetic energy and
obtain the MG phase \cite{giamarchi_commensurate_long}. Although
we have done the derivation of the Mott phase for fermions in one
dimension we expect its physics to survive into higher dimension.
This can be seen by looking at the atomic limit (zero hoping) of
an interacting fermionic system (in any dimension). If the
repulsion extends over at least one interparticle distance,
leading to small values of $K$, particle hole excitations are
lowered in energy by excitonic effects. For example for fermions
with spins with both an onsite $U$ and a nearest neighbor $V$ the
gap to add one particle is $\Delta = U/2$. On the other hand the
minimal particle-hole excitations  would be to have the particle
and hole on neighboring sites (excitons) and cost $\Delta_{\rm
p.h.}=U - V$, as shown in Figure~\ref{fig:exciton}.
When disorder is added the gaps decrease respectively
\cite{footnote_bounded_disorder} as $\Delta \to \Delta -W$ and
$\Delta_{\rm p.h.} \to \Delta_{\rm p.h.} - 2W$. Thus the
conductivity gap  closes, the compressibility remaining zero (for
bounded disorder). According to this physical picture of the MG,
the low frequency behavior of conductivity is dominated by
excitons (involving neighboring sites). This is at variance from
the AG where the particle and the hole are created on distant
sites. This may have consequences on the precise low frequency
form of the conductivity such as logarithmic corrections. When
hopping is restored, we expect the excitons to dissociate and the
MG to disappear above a critical value $K>K^*$. Since finite
range is needed for the interactions, in all cases (fermions or
bosons) $K^* < 1$. In addition we expect $K^* < 1/2$ for fermions
with spins. One interesting question is the question of d.c.
transport in the three phases, and whether the Mott glass has a
d.c. transport closer to the Anderson or the Mott phase. Since the
excitons are neutral, one simple guess would be that such
excitations would not contribute to the d.c. transport. The d.c.
conductivity in the Mott glass phase would thus be still
exponentially activated just as in the Mott insulator. Of course
more detailed studies would be needed to confirm this point.

\section{Conclusions}

Many disordered fermionic system can thus be successfully
described by an elastic disordered theory. In one dimension, this
situation is ubiquitous due to the importance of collective
excitations. Most physical system, whether one starts with
fermions, bosons or spins, can be represented in terms of bosonic
excitations. In higher dimension such a description is valid in
crystalline phases such as a Wigner crystal in which the quantum
particles are strongly localized due to their interactions or
charge density waves. Disorder then leads to rich physical
phenomena coming from the competition between the elasticity,
wanting a well ordered structure and the disorder that distorts
the structure to gain pinning energy. This leads to the existence
of many metastable configurations and to glassy properties. Using
the various methods described in these notes, we now have a good
description of the low energy excitations of such structures.
This gives access to a host of physical properties such as the
a.c. transport properties.

Clearly one of the most important open questions is the the issue
of topological defects in such structures. Indeed, such defects
are needed to describe the melting of these cristalline phases
and will be necessary to go to more ``liquid'' phases in which the
statistics (fermionic or bosonic) of the particles will play a
much more crucial role. In addition d.c. transport is obviously
dominated by such excitations. Unfortunately so far the methods
able to tackle the properties in the localized phase such as the
Gaussian Variational Method cannot handle such solitonic
excitations, so radically new methods will need to be designed to
handle them.

\section*{Acknowledgements}

The work presented in these notes results from many fruitful and
enjoyable collaborations. We would like to thank R. Chitra, H.
Maurey, B. S. Shastry and specially P. Le Doussal. Most
importantly, nothing would have started without an initial
collaboration with H. J. Schulz, to the memory of whom we would
like to dedicate these lecture notes.


\begin{thebibliography}{10}
\bibitem{rcra}
There are some subtleties for systems for which the size of the
particles
  differs from the lattice spacing. This leads to the definitions of two
  characteristic lengths. We ignore these differences here since they do not
  occur for classical CDW and 1D fermionic systems. See
  \cite{giamarchi_book_young} for consequences for classical systems and
  \cite{chitra_wigner_hall} for the Wigner crystal.
\bibitem{loflim_footnote}
Such a length was also obtained by Imry and Ma for the random
field Ising
  model. So a proper acronym for this length might be the Loflim length (for
  Larkin-Ovchinikov, Fukuyama-Lee, Imry-Ma).
\bibitem{footnote_bounded_disorder}
For small bounded disorder the gap of the Mott insulator phase is
robust.
\end{thebibliography}

\end{document}